\documentclass[a4paper,fleqn,usenatbib,useAMS]{mnras}
\usepackage[T1]{fontenc}
\usepackage{ae,aecompl}

\usepackage{graphicx}   
\usepackage{amsmath}    
\usepackage{amssymb}    
\usepackage{breakurl}
\usepackage{url}
\usepackage{subfig}
\usepackage{color}
\usepackage{verbatim} 
\usepackage[normalem]{ulem} 

\makeindex \citeindextrue

\newcommand{\rmu}{rad\,m$^{-2}$}
\newcommand{\Fermi}{\textit{Fermi}}

\title[Multi-wavelength study of S4~1030+61]{Multi-wavelength observations of the $\gamma$-ray flaring quasar S4~1030+61 in 2009-2014}
\author[E. V. Kravchenko et al.]{E. V. Kravchenko$^{1}$\thanks{Contact e-mail: \href{mailto:evgenia.v.kravchenko@gmail.com}{evgenia.v.kravchenko@gmail.com}}, 
Y. Y. Kovalev$^{1,2}$, T. Hovatta$^{3,4}$, V. Ramakrishnan$^{3}$\\
$^{1}$Lebedev Physical Institute, Astro Space Center, Profsoyuznaya 84/32, Moscow 117997, Russia\\
$^{2}$Max-Planck-Institut f\"ur Radioastronomie, Auf dem H\"ugel 69, Bonn D-53121, Germany\\
$^{3}$Aalto University Mets\"ahovi Radio Observatory, Mets\"ahovintie 114, FI-02540 Kylm\"al\"a, Finland\\
$^{4}$Aalto University Department of Radio Science and Engineering, P.O. BOX 13000, FI-00076 Aalto, Finland\\
}
\date{Accepted 2016 July 19. Received 2016 July 19; in original form 2016 February 21}
\pubyear{2016}

\begin{document}
\label{firstpage}
\pagerange{\pageref{firstpage}--\pageref{lastpage}}
\maketitle

\begin{abstract}
We present a study of the parsec-scale multi-frequency properties of the quasar S4~1030+61 during a prolonged radio and $\gamma$-ray activity.
Observations were performed within \textit{Fermi} $\gamma$-ray telescope, OVRO 40-m telescope and MOJAVE VLBA monitoring programs, covering five years from 2009.
The data are supplemented by four-epoch VLBA observations at 5, 8, 15, 24, and 43~GHz, which were triggered by the bright $\gamma$-ray flare, registered in the quasar in 2010.
The S4~1030+61 jet exhibits an apparent superluminal velocity of (6.4$\pm$0.4)c and does not show ejections of new components in the observed period, while decomposition of the radio light curve reveals nine prominent flares. 
The measured variability parameters of the source show values typical for \textit{Fermi}-detected quasars.
Combined analysis of radio and $\gamma$-ray emission implies a spatial separation between emitting regions at these bands of about 12~pc and locates the $\gamma$-ray emission within a parsec from the central engine.
We detected changes in the value and direction of the linear polarization and the Faraday rotation measure. 
The value of the intrinsic brightness temperature of the core is above the equipartition state, while its value as a function of distance from the core is well approximated by the power-law.
Altogether these results show that the radio flaring activity of the quasar is accompanied by injection of relativistic particles and energy losses at the jet base, while S4~1030+61 has a stable, straight jet well described by standard conical jet theories.
\end{abstract}


\section{Introduction}

Multi-wavelength observations are a powerful tool to study the emission mechanisms at subparsec and parsec-scales of active galactic nuclei (AGN). 
Relativistic jets of quasars are oriented close to the line of sight, which in combination with the Doppler boosting effects makes them to be some of the most powerful objects in the Universe and to show strong variability at wavelengths from radio to $\gamma$ rays.

Whereas the very long baseline radio interferometry (VLBI) technique enables us to study subparsec-to-parsec structure of quasars, $\gamma$-ray telescopes have much poorer resolution and the location of this emission is not established well.

\begin{figure*}
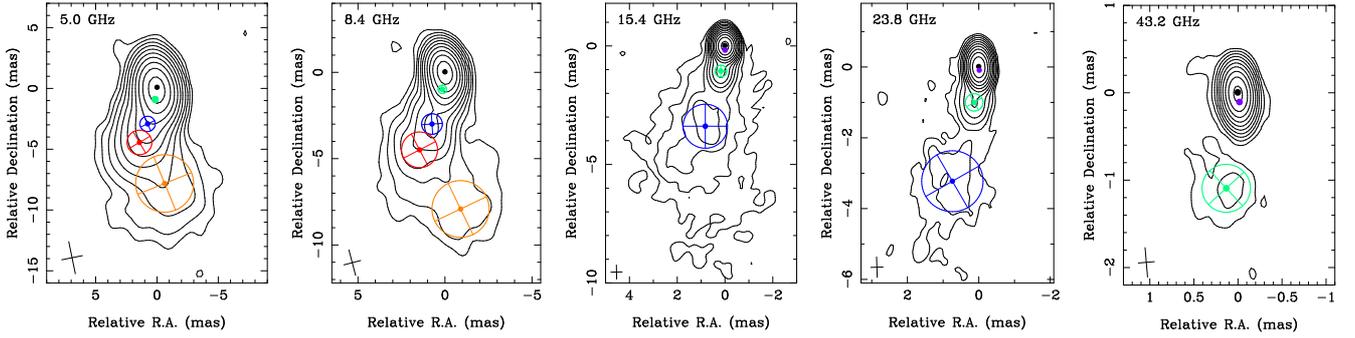

\centering
\includegraphics[angle=-90,scale=.26]{1030+611_C2_Istack1.eps}\quad
\includegraphics[angle=-90,scale=.26]{1030+611_X1_Istack1.eps}\quad
\includegraphics[angle=-90,scale=.26]{1030+611_U1_Istack1.eps}\quad
\includegraphics[angle=-90,scale=.26]{1030+611_K1_Istack1.eps}\quad
\includegraphics[angle=-90,scale=.26]{1030+611_Q1_Istack1.eps}\\
\caption{Stacked naturally weighted contour images of total intensity over four multi-wavelength epochs 
at 5.0, 8.4, 15.4, 23.8 and 43.2~GHz (from left to right).
Contours of equal intensity are plotted starting from 3~rms level at $\times$2 steps.
The 1 rms value equals to 0.07, 0.07, 0.14, 0.10 and 0.23~mJy at 5.0, 8.1, 15.4, 23.8 and 43.2~GHz respectively.
The full width at half maximum (FWHM) of the synthesized beam is shown in the bottom left corner of every image.
Color circles represent model-fit components.
The color figures are available in the electronic version of the article.}
\label{fig_cxust}
\end{figure*}

\begin{figure*}
\centering
\includegraphics[angle=-90,scale=.7]{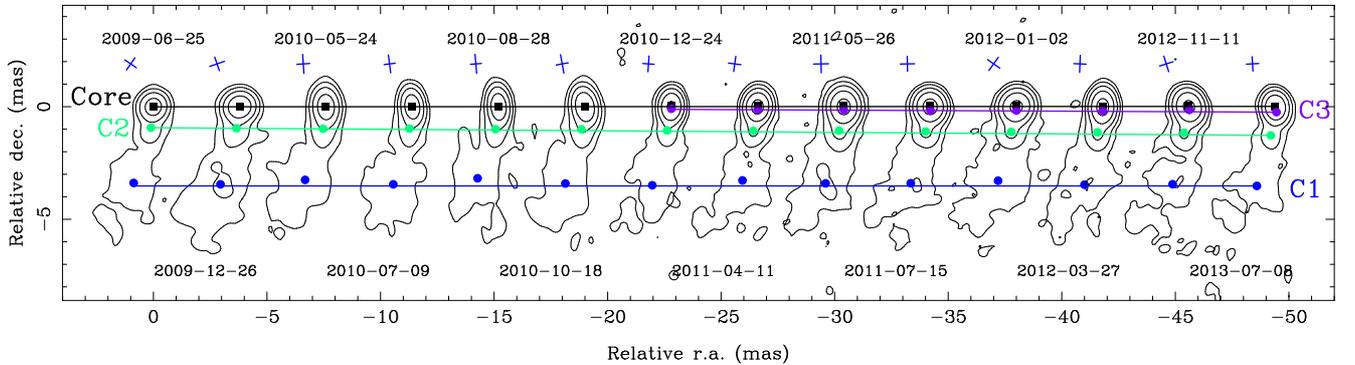}
\caption{S4~1030+61 naturally weighted contour images of total intensity at 15.4~GHz from 2009 (left) to 2013 (right) for all the MOJAVE epochs as listed in Sections~\ref{s:multifreqVLBA} and \ref{s:MOJAVE}.
Contours of equal intensity are plotted starting from 4 r.m.s.\ level at $\times$5 steps. 
FWHM of the synthesized beam is shown by crosses.
Symbols represent model-fit components, while solid lines connect the positions of components at first and last epochs.
The designation of the components is given in Section~\ref{s:struct}.}
\label{fig_uall}
\end{figure*}

Many authors have discussed the connection between radio and $\gamma$-emission.
\citet{vt_96,jorstad_etal01,lv_03,kovalev_etal09,pushkarev_etal10,leon-tavares_etal11,fuhrmann_etal14,ramakrishnan_etal15} and \citet{karamanavis_etal16} reported significant correlations between these bands. 
While localizing the $\gamma$-ray emitting region authors agree that it originates in the relativistic jet of AGN, but dispute about more precise location of this emission within a jet. 
Two scenarios are favoured: the sub-parsec region around the central massive black hole \citep[e.g.,][]{bl_95,tavecchio_etal10,dotson_etal2012} and the regions located few parsecs downstream the jet, where jet becomes transparent at radio wavelengths \citep[e.g.,][]{lv_03}. 
The parsec-scale core was identified by \cite{kovalev_etal09,pushkarev_etal10} as a likely location for both the $\gamma$-ray and radio flares, which appears within typical timescales of up to a few months of each other. 
Many $\gamma$-ray flares are connected with the ejection of a newly born components \citep[e.g.,][]{jorstad_etal01,agudo_etal11} but sometimes no observed components can be associated with the $\gamma$-ray flares \citep[e.g.,][]{pk_98}, though many sources show dual properties \citep[e.g.,][]{marscher_etal08,ramakrishnan_etal14,morozova_etal14}.
\begin{table}
\caption{VLBA central frequencies. The full table is available online.\label{tb_frq}}
\begin{tabular}{lcr}
\hline
IEEE band & IF & Frequency\\
 & &(MHz)\\
\hline
C & 1 & 4604.5\\
  & 2 & 4612.5\\
  & 3 & 4999.9\\
  & 4 & 5007.5\\
X & 1 & 8104.5\\
  & 2 & 8112.5\\
\hline
\end{tabular}
\end{table}

\begin{table}
 \caption{Amplitude corrections for the S2087E VLBA experiment. The full table is available online.\label{tb_acorr}}
  \begin{tabular}{@{}cccccc@{}}
\hline
Antenna & Band & Epoch & IF & Polarization & Correction\\
\hline
BR & C & All & 1,2 & RCP & 1.10\\
KP & C & All & 1,2 & RCP & 1.10\\
KP & C & All & 1,2 & LCP & 1.13\\ 
MK & C & All & 1 & LCP & 1.10\\
PT & C & All & 4 & RCP & 0.90\\
FD & X & All & 1,2 & RCP & 1.08\\
\hline
\end{tabular}
\end{table}

In this paper we present multi-epoch multi-frequency VLBI analysis of the flat spectrum radio quasar S4~1030+61 (TXS 1030+611, 1FGL J1033.8+6048), which shows strong variability at radio and $\gamma$-ray wavelengths. 
This is a high optical polarization quasar at a redshift of 1.4009  \citep{schneider_etal10,2012AJ....143..119I}. 
It is being monitored by the Large Area Telescope (LAT) on board of \textit{Fermi Gamma-ray Space Telescope} (\Fermi) and by the 40-m radio telescope of the Owens Valley Radio Observatory (OVRO) since 2008, providing good time coverage. 
After a bright $\gamma$-ray flare occurred in 2010~\citep[][]{ciprini_atel10,smith_atel10,carrasco_atel10}, a four-epoch VLBA campaign at 5, 8, 15, 24, and 43~GHz was started. These VLBI observations are supplemented by observations within the MOJAVE\footnote{Stands for Monitoring Of Jets in Active galactic nuclei with VLBA experiments~\citep[][]{lister_etal09,lister_etal13}} program made in 2009---2013. Here we report a detailed kinematic, polarized and radio-to-$\gamma$-ray study of the blazar.

\begin{figure}
\centering
\includegraphics[scale=.31]{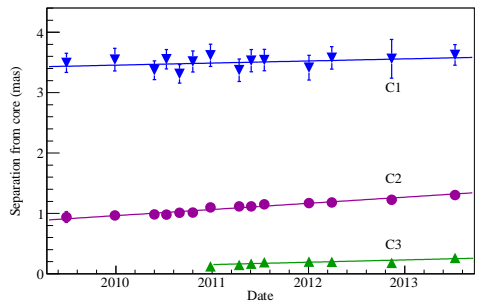}
\caption{Angular separation of the modelled components from the core versus time at 15.4~GHz.
The solid lines are a weighted linear fit to the estimated core distances.
The accuracy in position is estimated with equation~(\ref{eq:posacc}). The error bars of C2 and C3 component positions are smaller than the symbol size.}
\label{fig_kins}
\end{figure}
\begin{figure}
\centering
\includegraphics[scale=.33]{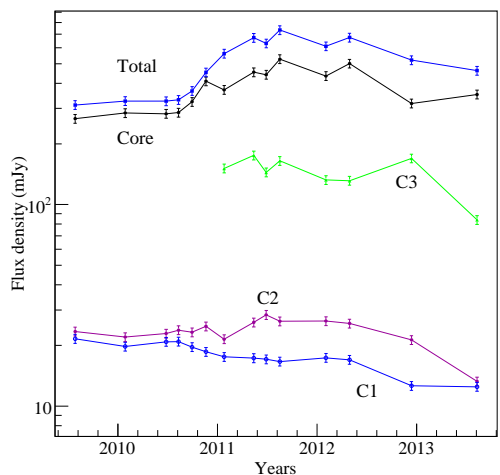}
\caption{Radio light curves of the modelled components at 15.4~GHz for 14 observational VLBA epochs.}
\label{fig_kin}
\end{figure}

This paper is structured as follows: in Section 2 we describe VLBA, OVRO, and \Fermi\ observations, and data reduction. In Section 3 we present results of the jet structure and kinematics study, core shift analysis, study of the polarization properties, decomposition of the radio light curve onto flares, and joint radio-$\gamma$-ray analysis. In Section 4 we discuss these results, with the astrophysical interpretation touched. Section 5 summarizes our findings. Through the paper the model of flat $\Lambda$CDM cosmology was used with $H_0$ = 68~km s$^{-1}$, $\Omega_M$ = 0.3, and $\Omega_{\Lambda}$ = 0.7 \citep[][]{planck_15}. 
The angular size of 1~mas in this cosmology corresponds to 8.68~pc linear scale at the redshift 1.4009. 
The luminosity distance is 10320.2~Mpc.
We refer the `radio core` at each frequency as the brightest component at the apparent, stationary base of the jet and associate it with the region of an optical depth $\tau \thickapprox 1$ \citep{marscher08}.
The position angles (PA) are measured from North through East.
The spectral index $\alpha$ is defined as $S \propto \nu^{\alpha}$, where $S$ is the flux density, observed at frequency $\nu$.

\begin{table*}
  \caption{Results of Gaussian model fitting and component parameters at 4.6--43.2~GHz. Column designation: (1) observation date; (2) the name of the component ("U" stands for unclassified); (3) the integrated flux density in the component and its error; (4) the radial distance of the component center from the center of the image and its error; (5) the position angle of the center of the component relative to the image center; (6) the FWHM major axis of the fitted Gaussian; (7) axial ratio of major and minor axes of fitted Gaussian; (8) major axis position angle of fitted Gaussian. The full table is available online.\label{tb_mdf}}
  \begin{tabular}{@{}lcr@{$\pm$}lr@{$\pm$}lcccc@{}}
\hline
Date & Name & \multicolumn{2}{c}{Flux density} & \multicolumn{2}{c}{Distance} & P.A. & Major & Ratio & Major P.A.\\
 &  & \multicolumn{2}{c}{(Jy)} & \multicolumn{2}{c}{(mas)} &($^\circ$)& (mas) & & ($^\circ$) \\
(1)& (2) & \multicolumn{2}{c}{(3)} & \multicolumn{2}{c}{(4)} &(5)& (6) &(7) & (8) \\
\hline
\multicolumn{10}{c}{4.6~GHz}\\
\hline
2010--05--24
& Core & 0.147&0.007&0.00&0.04&0.0&0.202&1.0&\dots\\
& C2 & 0.055&0.003&1.02&0.07&173.8&0.480&1.0&\dots\\
& C1 & 0.0263&0.0013&3.02&0.14&166.0&1.094&1.0&\dots\\
& U1 & 0.0210&0.0011&4.43&0.18&162.9&1.918&1.0&\dots\\
& U0 & 0.0126&0.0006&7.5&0.4&$-$177.6&4.607&1.0&\dots\\
2010--07--09
& Core & 0.147&0.007&0.00&0.04&0.0&0.146&1.0&\dots\\
 \hline
\end{tabular}
\end{table*}
\begin{table*}
\begin{center}
\caption{Measured integrated flux density (I), linear polarization (p) and degree of polarization (m), electric vector position angle and rms noises at the central region (corresponds to the peak flux density of the map) of S4~1030+61 maps at seven frequencies.\label{tb_fp}}
\begin{tabular}{ccccccccc}
\hline
Parameter & Epoch$^a$ & \multicolumn{7}{c}{Frequency (GHz)}\\
\cline{3-9}
 &  & 4.6 & 5.0 & 8.1 & 8.4 & 15.4 & 23.8 & 43.2\\
\hline
I (mJy) & 1 & 183.4 & 183.4 & 219.4 & 228.3 & 279.1 & 329.5 & 401.5\\
  & 2 & 185.4 & 184.5 & 216.4 & 223.9 & 282.8 & 365.8 & 470.3\\
  & 3 & 210.1 & 209.2 & 238.0 & 243.7 & 320.1 & 452.0 & 628.7\\
  & 4 & 220.2 & 216.1 & 271.4 & 278.5 & 404.4 & 585.6 & 757.2\\
p (mJy) & 1 & 2.5 & 2.9 & 6.6 & 6.7 & 3.3 & 1.7 & 7.8\\
  & 2 & 4.9 & 5.2 & 8.6 & 8.9 & 8.6 & 6.2 & 5.1\\
  & 3 & 5.0 & 4.9 & 9.4 & 10.0& 7.7 & 4.8 & 6.9\\
  & 4 & 4.2 & 3.7 & 8.8 & 9.5 & 5.1 & 3.6 & 7.4\\
m (\%) & 1 & 1.4 & 1.6 & 3.0 & 2.94& 1.2 & 0.5 & 1.9\\
  & 2 & 2.6 & 2.8 & 4.0 & 4.0 & 3.0 & 1.7 & 1.1\\
  & 3 & 2.4 & 2.3 & 4.0 & 4.1 & 2.4 & 1.1 & 1.1\\
  & 4 & 1.9 & 1.7 & 3.2 & 3.4 & 1.3 & 0.6 & 1.0\\
EVPA (deg) & 1 & 86.5 & 81.7 & 54.6 & 50.3 & 49.1 & 74.0 & 165.4\\
           & 2 & 76.9 & 66.5 & 55.9 & 53.3 & 68.3 & 61.3 & 34.5\\
           & 3 & 74.3 & 66.5 & 60.3 & 57.5 & 69.4 & 80.5 & 64.1\\
           & 4 & 80.2 & 76.0 & 73.0 & 70.5 & 96.7 & 108.7 & 57.7\\
$\sigma_I$ (mJy) & 1 & 0.15 & 0.14 & 0.15 & 0.4 & 0.18 & 0.17 & 0.4\\
           & 2 & 0.18 & 0.12 & 0.14 & 0.12 & 0.17 & 0.2 & 0.6\\
           & 3 & 0.16 & 0.14 & 0.14 & 0.14 & 0.17 & 0.2 & 0.5\\
           & 4 & 0.2 & 0.13 & 0.12 & 0.12 & 0.16 & 0.19 & 0.3\\
$\sigma_p$ (mJy) & 1 & 0.2 & 0.15 & 0.16 & 0.17 & 0.17 & 0.21 & 0.5\\
	   & 2 & 0.18 & 0.15 & 0.16 & 0.15 & 0.18 & 0.3 & 0.6\\
	   & 3 & 0.19 & 0.16 & 0.16 & 0.15 & 0.18 & 0.3 & 0.5\\
	   & 4 & 0.2 & 0.16 & 0.16 & 0.14 & 0.17 & 0.2 & 0.4\\
\hline
\multicolumn{9}{l}{$^a$ Epochs are labeled as follows: 
1 for 2010--05--24, 
2 for 2010--07--09, 
3 for 2010--08--28, 
4 for 2010--10--18.}\\
\end{tabular}
\end{center}
\end{table*}

\section{Observations and data reduction}
\subsection{Multi-epoch 4--43~GHz VLBA observations}
\label{s:multifreqVLBA}

The source S4~1030+61 was observed (code S2087E) with the Very Long Baseline Array (VLBA) of the National Radio Astronomy Observatory (NRAO) during four sessions: 2010--05--24, 2010--07--09, 2010--08--28, and 2010--10--18 (noted as `epochs' below).
Full configuration of the array was used at C, X, K$_u$, K and Q frequency bands (according to the IEEE nomenclature), that corresponds to 6, 4, 2, 1.3, and 0.7~cm wavelengths respectively (see Table~\ref{tb_frq}). 
In each band four 8~MHz-wide frequency channels (IFs) were recorded in right and left circular polarization with 2-bit sampling and total recording rate of 256~Mbps. 
The data were correlated at the NRAO VLBA Operations Center in Socorro (New Mexico, USA) with averaging time of 2 seconds.
C and X bands are split into two sub-bands (each of 16~MHz width) centered at 4608.5, 5003.5, 8108.5, and 8429.5 MHz respectively and in the following analysis were processed independently.
K$_u$, K and Q bands were not split into sub-bands, resulting in 32~MHz band width and centered at 15365.5, 23804.5 and 43217.5~MHz.
On-source time at each epoch is of about 6~hr equally distributed over all frequencies.

The data were processed within the NRAO Astronomical Image Processing System \citep[AIPS,][]{aips} with usage of the Caltech Difmap \citep{shepherd_etal94,shepherd_97}, following the standard procedure\footnote{\url{http://www.aips.nrao.edu/cook.html}}.
It includes i) a~priori amplitude calibration with measured antenna gain curves and system temperatures; 
ii) phase calibration using the pulse phase calibration signal injected during the observations; 
iii) removal of residual phase delay and delay rate (known as `fringe fitting`). 
During the fringe fitting a separate solution for the group delay and phase rate was made for each sub-band.
The fringe-fit interval is chosen to be 2 minutes for C-, X-, and K$_\mathrm{u}$-bands, 1 minute for K-band, and 30 seconds for Q-band. 
After fringe fitting, a complex bandpass calibration was made and residual amplitude corrections determined and applied, using AIPS task LPCAL, for a few sub-bands at specific antennas (see Table~\ref{tb_acorr}).
The estimated accuracy of the VLBA amplitude calibration in the 5-15~GHz frequency range is of $\thicksim$5\% and at 15-43~GHz of $\thicksim$10\% \citep[see also][]{kovalev_etal05,sokolovsky_etal11}.

\begin{table}
\caption{Kinematics of S4~1030+61 jet components.\label{tb_kin}}
\begin{center}
\begin{tabular}{cccc}
\hline
Component & ${\beta}_\mathrm{app}$ & Proper motion & Ejection epoch\\
 ID & ($c$) & ($\mu$as/yr) & (yr)\\
\hline
C1 & 2$\pm$2 & 31$\pm$24  & 1899$\pm$85\\
C2 & 6.4$\pm$0.4 & 97$\pm$6 & 2000.0$\pm$0.7\\
C3 & 2.7$\pm$0.7 & 40$\pm$10 & 2007.4$\pm$1.2\\
\hline
\end{tabular}
\end{center}
\end{table}

Polarization leakage parameters of the antennas (D-terms) 
were determined within AIPS with task LPCAL \citep{1995AJ....110.2479L}. 
D-terms were examined for time stability \citep{roberts_etal94,gomez_etal02} using the two polarizations of each IF at all antennas. 
Corrections have been calculated at C, X, K, and Q bands by comparing the R-L offset over one year time interval (VLBA experiments s2087 and bk134); and at K$_u$ band by connecting our observations with the D-term phase solutions from 10-year MOJAVE VLBA program data.
Thus, accuracy of leakage parameters approaches 2\degr\ at C and X, 1\degr\ at K$_u$, and 3\degr\ at K and Q bands. 

Absolute EVPA calibration was done using the position angles of 0851$+$202 (OJ~287) and 0923$+$392 from the VLA/VLBA Monitoring Program\footnote{\url{http://www.vla.nrao.edu/astro/calib/polar/}} \citep{taylor_myers_00}, 
the University of Michigan Radio Astronomy Observatory monitoring program\footnote{\url{https://dept.astro.lsa.umich.edu/datasets/umrao.php}}, and from the MOJAVE\footnote{\url{http://www.physics.purdue.edu/astro/MOJAVE/index.html}} VLBA program. 
Absolute calibration error is assessed as the difference in corrected EVPA between pairs of sub-bands for C and X bands, 
and at K$_u$, K and Q band it equals deviation of corrected EVPA from the EVPA of calibrators taken from the monitoring programs.
The resulting absolute EVPA calibration error 2\degr\ for the K$_u$ band, and 4\degr\ for other frequencies.
Final calibration error of absolute vector position angle comprises errors from instrumental and 
absolute calibrations and is estimated to be 4.5\degr\ at low frequencies (C and X), 
2.5\degr\ at K$_u$ and 5\degr\ at K and Q bands.

\subsection{Multi-epoch 15~GHz MOJAVE observations}
\label{s:MOJAVE}
We supplemented our analysis with the data obtained within the MOJAVE program. 
Observations are done at 15.4 GHz with VLBA at ten epochs: 
2009--06--25, 2009--12--26, 2010--12--24, 2011--04--11, 2011--05--26, 2011--07--15, 2012--01-02,
2012--03--27, 2012--11--11, and 2013--07--08.
The fully calibrated publicly available data were used.
Details on the data reduction and calibration are given in ~\cite{lister_etal09}.
The absolute flux density of the observations is accurate within 5\%~\citep{lister_homan_05}. The average uncertainty in EVPA amounts to 5\degr~\citep{lister_homan_05,hovatta_etal12}.

\subsection{15~GHz OVRO observations}

Public data\footnote{\url{http://www.astro.caltech.edu/ovroblazars/}} of S4~1030$+$61 observations within the Owens Valley Radio Observatory 40-m Telescope monitoring program \citep{richards_etal11} were used in the analysis.
Observations are done at 15~GHz in a 3~GHz bandwidth from 2008--06--20 to 2014--01--21 about twice per week.
Details of the data reduction and calibration are given in ~\cite{richards_etal11}.

\subsection{Gamma-ray \textit{Fermi}/LAT data}

The $\gamma$-ray fluxes in the range 0.1-200 GeV were obtained with the LAT onboard the space \Fermi\ $\gamma$-ray observatory from 2008--08--04 to 2014--02--23. 
The analysis of the data was done using the \Fermi\ ScienceTools software package\footnote{\url{http://fermi.gsfc.nasa.gov/ssc/data/analysis/documentation/Cicerone/}}
version v10r0p5. 
We used Pass 8 data and the instrument response function P8R2\_SOURCE\_V6. The diffuse models gll\_iem\_v06.fits and iso\_P8R2\_SOURCE\_V6\_v06.txt were used.
The full light curve was built using 1~week time binning. 
The sources from 3FGL LAT catalog~\citep{acero_etal15} with the high test statistics \citep[TS, ][]{Mattox_etal96} were used to generate the source model.
The source was modelled using a log-parabola ($dN/dE=N_0(E/E_b)^{-(\alpha+\beta\mathrm{log}(E/E_b))}$, where $\alpha$=2.12 and $\beta$=0.11), remaining constant during full observed period. 
For each time bin, the integrated flux values were computed using the maximum-likelihood algorithm implemented in the science tool \textit{gtlike} to estimate significance of the results. 
We used a detection criteria which corresponds to a maximum-likelihood test statistic TS $>4$ or the number of model predicted photons is higher than 10 \citep{abdo_etal11}.
Otherwise a 2$\sigma$ upper limit of the flux was computed.

\subsection{Data fitting algorithm}

To model structure of the source by a number of circular/elliptical two-dimensional Gaussian components were fitted to the fully calibrated visibility data (in the $uv$ plane), using task \textit{modelfit} in DIFMAP.

The MINUIT\footnote{Numerical function minimization and error analysis}~\citep{minuit} algorithm, implemented within the ROOT\footnote{Stands for Rapid Object-Oriented Technology}~\citep{root} framework, was used for the other fitting procedures through the paper.
It derives parameters of a fit using the method of maximum likelihood by searching for the global minimum of a multi-parametric function.
The width of the function minimum, or the shape of the function in some neighbourhood of the minimum, gives information about the uncertainty in the best parameter values, taking into account correlations between parameters.
For the linear models MINUIT turns into usual least squares analysis.

\begin{figure*}
\centering
\includegraphics[scale=.76]{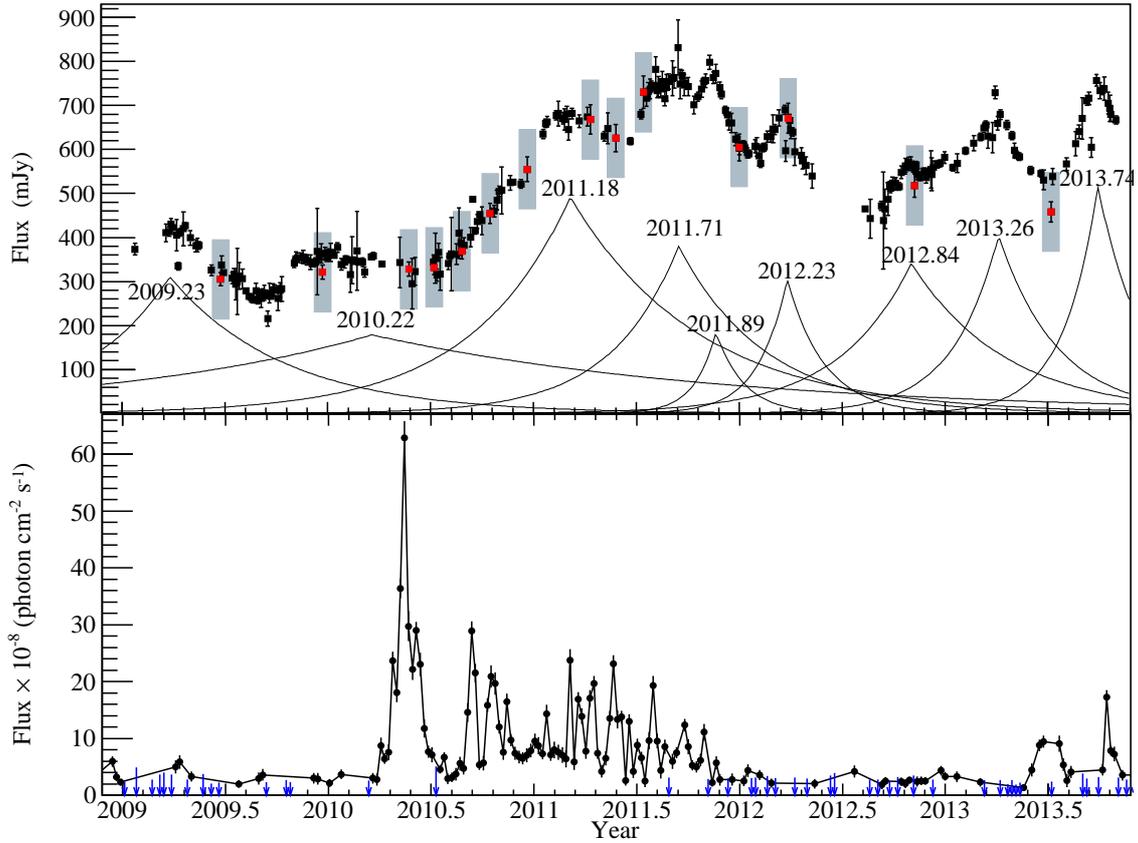}
\caption{(top) Decomposed OVRO 15~GHz flux density curve (points) into exponential flares (solid lines). Filled rectangles together with red points indicate VLBA total flux density at 15.4~GHz, observed within MOJAVE program. The epochs of the flare maximum are labeled above each one of them.
(bottom) \textit{Fermi} weekly-binned $\gamma$-ray light curve at 0.1--200~GeV. Upper limits are given by blue arrows.}
\label{fig_ovro}
\end{figure*}

\section{Results}
\subsection{Jet structure}
\label{s:struct}
S4~1030+61 has a one-sided jet structure, apparently short at mm and extending up to 13 mas at cm wavelengths.
Stacked images of the source at 5.0, 8.1, 15.4, 23.8 and 43.2 GHz are shown in Fig.~\ref{fig_cxust}.
The jet position angle at sub-mas scale is almost 180\degr, meanwhile at mas scales changes to $-$170\degr.
The low-frequency image (Fig.~\ref{fig_cxust}) shows that the jet bends at $\thicksim5$ mas from the core.
The structure of the source is modelled to consist of five at 4.6 to 8.4~GHz, four at 15.4 and 23.8~GHz, and three Gaussian components at 43.2~GHz (see Table~\ref{tb_mdf}).
All jet components have circular shape, except for the core in few cases discussed below.

In the 4.6--8.4~GHz range the core component is modelled to be elliptical through all epochs.
At 15.4~GHz the core has elliptical shape extended toward a direction of the jet from 2009--12--26 to 2010--10--18. 
Starting from 2010--12--24, the source model became more reliable by introducing an additional component apart from the core (see discussion below). 
From this date, the core could be also modelled with a circularly-shaped Gaussian component. 
Naturally weighted CLEAN images at 15.4 GHz are shown in Fig.~\ref{fig_uall}.
At higher frequencies we model this new component through all four epochs separately from the core. 
At 43.2~GHz the emission in the extended jet region of the source is resolved, and model contains three components.

Theoretical models of AGN jets predict elongated shape of the core component \citep[e.g.,][]{blandford_koenigl_79,1988ApJ...334..539D}, arising from the conical or cylindrical jet geometry. Moreover, the VLBI studies of the jet samples indicate that core components are well described by elliptical Gaussians extended along the jet direction \citep[e.g.,][]{kovalev_etal05,lee_etal08}.
We have analyzed if the data are significantly better described if the new C3 component is introduced instead of explaining changes in the structure by a varying core component extension and flux density. The following two models of the source at 15.4 GHz have been considered: (i) the central feature (core and C3) has been modelled by a single elliptically-shaped Gaussian component, other jet component are circular, (ii) core, C3 and other jet components are modelled as separate circular Gaussians. Before the 2010--12--24 the (i) and (ii) models result in comparable values of the reduced $\chi^2$ in the visibility domain and RMS intensity in the residual map. After this date, the model (ii) of the source structure is preferred by the both measures. This represents structural changes in S4 1030+61 core with time implying its circular shape and consequently prefers C3 as an individual component after 2010--12--24.

The components were visually cross-identified between frequencies on the basis of their position with respect to the core. 
Only three components were found to have consistent cross-identification through all frequencies. 
Their positions and designations are indicated in Fig.~\ref{fig_cxust} and Fig.~\ref{fig_uall}. 
The numeration of the components starts from the more distant component (C1) and increases (C2 and C3) towards the core.
Thus it can be seen, that the core and C3 components are blended at 4.6--15.4~GHz, resulting in elliptical shape of the modelled Gaussian core component.

\subsection{Kinematics}

During the analysis the core is assumed to be stationary over all the epochs.
The angular separation of the jet components from the core with time is shown on Fig.~\ref{fig_kins}.
The lower limit on accuracy in position of the modelled components in the image, $\sigma_\mathrm{position}$, can be estimated as \citep{fomalont_99,lee_etal08}:
\begin{equation}
\sigma_\mathrm{position} = {\frac{\theta_\mathrm{c}}{2S/N}}\sqrt{1+{\frac{S}{N}}}
\label{eq:posacc}
\end{equation}
mas, where the FWHM of the modelled size component $\theta_c$ and beam size $\theta_\mathrm{beam}$ is used, and the signal to noise ratio S/N is calculated using the peak flux density ($S$) of the component and the residual noise ($N$) of the image after subtraction of the model at the position of the component.
If the the component is unresolved, then the upper limit can be used to estimate component's size as \citep{lobanov_05,kovalev_etal05}
\begin{equation}
\theta_\mathrm{lim} = \theta_\mathrm{beam}\sqrt{\frac{4\mathrm{ln}2}{\pi}\mathrm{ln}{\frac{S/N}{S/N-1}}},
\label{eq:limitsize}
\end{equation}
where $\theta_\mathrm{beam}$ is a beam size.
The motion of the components was determined by least-square technique using linear dependence. 
All components meet good linear fit with no prominence of acceleration. 
Results of the fits are given in Table~\ref{tb_kin}, where transition from proper motion ($\mu$) to the apparent speed ($\beta_\mathrm{app}$) is done by
\begin{equation}
\beta_\mathrm{app} = \mu\ {\frac{D_\mathrm{L}}{(1+z)}}c,
\end{equation}
where $D_\mathrm{L}$ is the luminosity distance and $c$ is the speed of light.
Apparent velocities are measured to be $(2\pm2)c$ for the component C1, $(6.4\pm0.4)c$ for C2, and $(2.7\pm0.7)c$ for C3. 
The estimated ejection time, which corresponds to the components crossing the position of the 15~GHz core, is given in Table~\ref{tb_kin}.

Component C1 shows slow proper motion (see Fig.~\ref{fig_kins}) and its modelled size does not change significantly during the observed period, and it can be considered stationary.

\subsection{Flux density evolution}

The 15.4~GHz light curve of the modelled components is given in Fig.~\ref{fig_kin}.
The core flux density shows high variability during the observed period even after the separation of C3 component, and gives main contribution to the total flux density of the source.
Table~\ref{tb_fp} summarizes the information about S4~1030+61 fluxes at the seven frequencies and four epochs.

We combine data at 14 epochs made within the MOJAVE program with the four years of OVRO data. 
Superimposed light curves are shown on Fig.~\ref{fig_ovro}, where epochs from 2010--05--24 to 2010--10--18, are observed within our multi-wavelength campaign.
The value of VLBA total flux density is in good agreement with the single-dish OVRO flux density.
Therefore, there is no extended emission resolved at VLBI scales.
As seen in Fig.~\ref{fig_ovro} MOJAVE observations represent different activity states of the source.

\subsection{Evolution of the spectrum}
\label{s:spi}
\begin{figure}
\centering
\includegraphics[scale=.31]{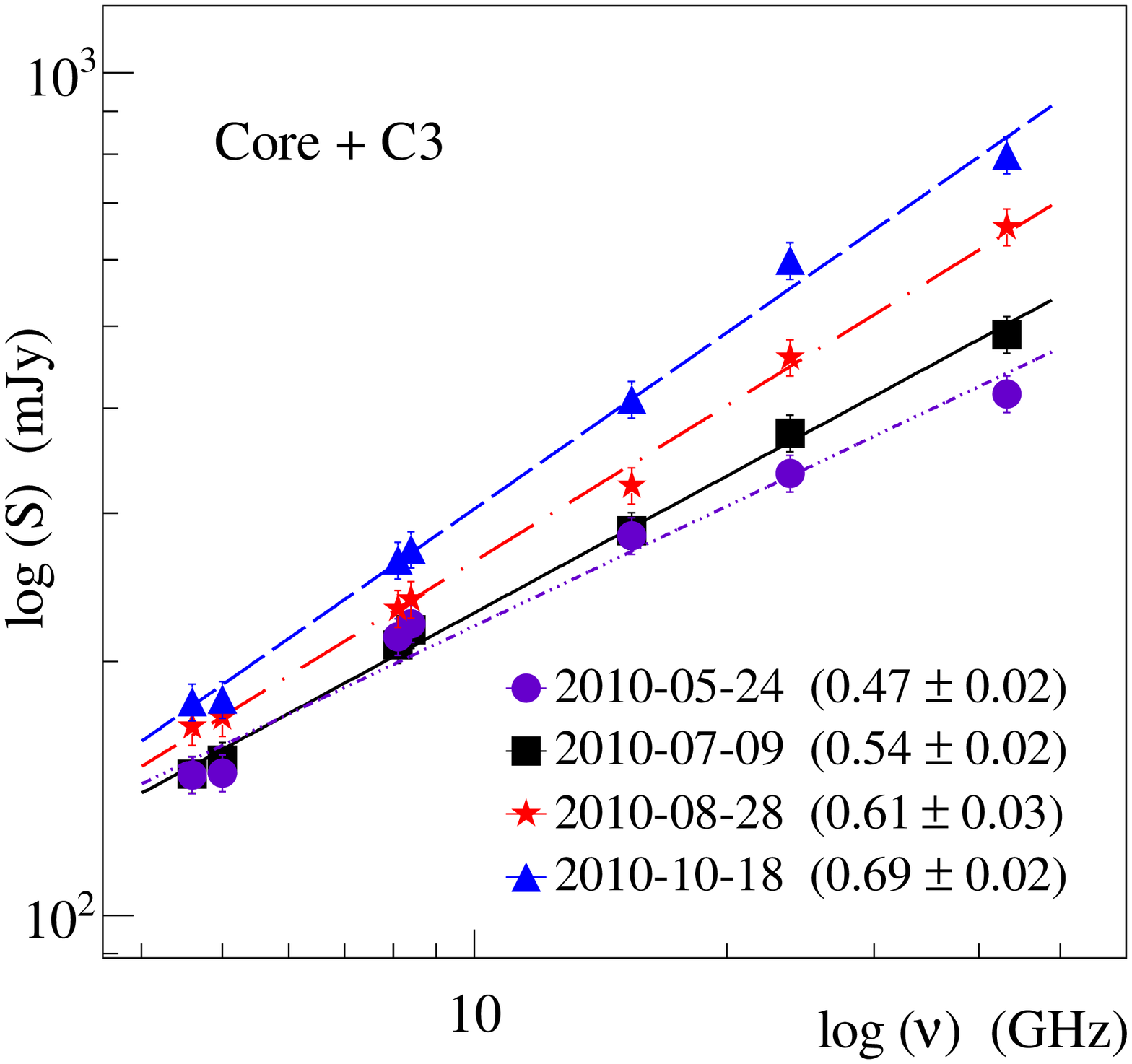}
\includegraphics[scale=.31]{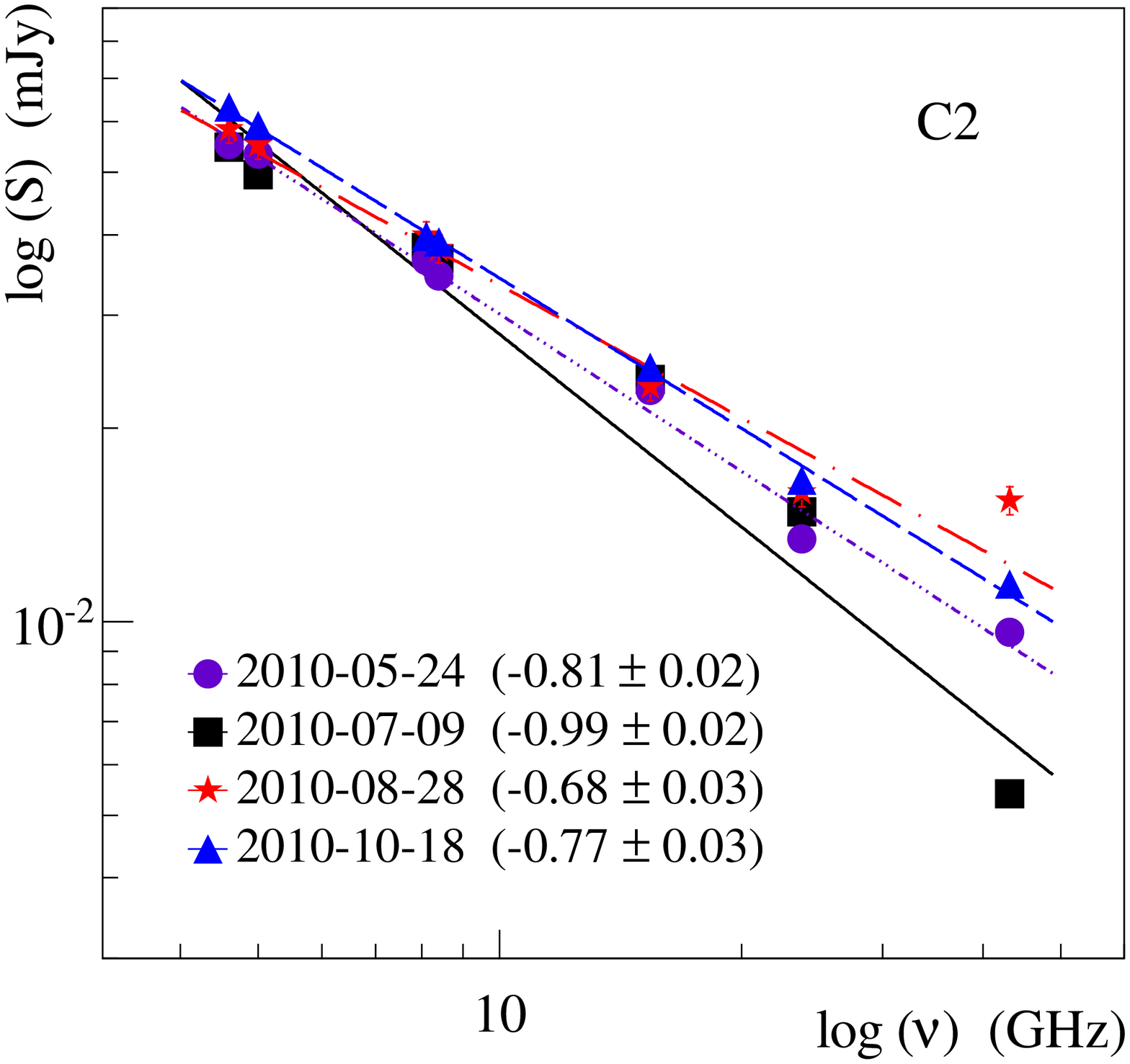}
\caption{Spectra of the core plus C3 (top) and C2 (bottom) components (symbols) together with the power-law fit results (lines) at four multiwavelength epochs. The fitted spectral indices are given in brackets. }
\label{fig_spi}
\end{figure}
The S4~1030$+$61 core region has a highly inverted spectrum, as shown in Fig.~\ref{fig_spi}.
The spectral index evolves in time from 0.47$\pm$0.02 at 2010--05--24 to 0.69$\pm$0.02 at 2010--10--18, as a flare rises.
This can be due to the passage of the C3 component through the core, which is modelled at 23.8--43.2~GHz separately, but blended with the core component at lower frequencies.
But most probably, hardening of the spectra can be explained by particle injection \citep{fromm_etal13} during the flare and energy losses~\citep{marscher_gear_85}.
Emission downstream the core shows optically thin emission (see example in Fig.~\ref{fig_spi}) with no signs of temporal evolution.

\subsection{Brightness temperature}
\label{s:tbr}
\begin{figure}
\centering
\includegraphics[scale=.36]{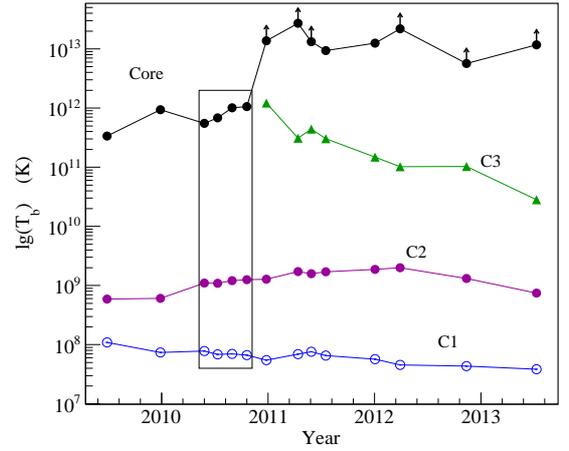}
\caption{Time evolution of the brightness temperature of the modelled components at 15.4~GHz at 14 VLBA epochs.
The black rectangle shows four epochs of a multiwavelength monitoring campaign. The arrows show upper limits, when the components are unresolved, calculated with equation~(\ref{eq:limitsize}).}
\label{fig_tb}
\end{figure}

\begin{figure}
\centering
\includegraphics[scale=.34]{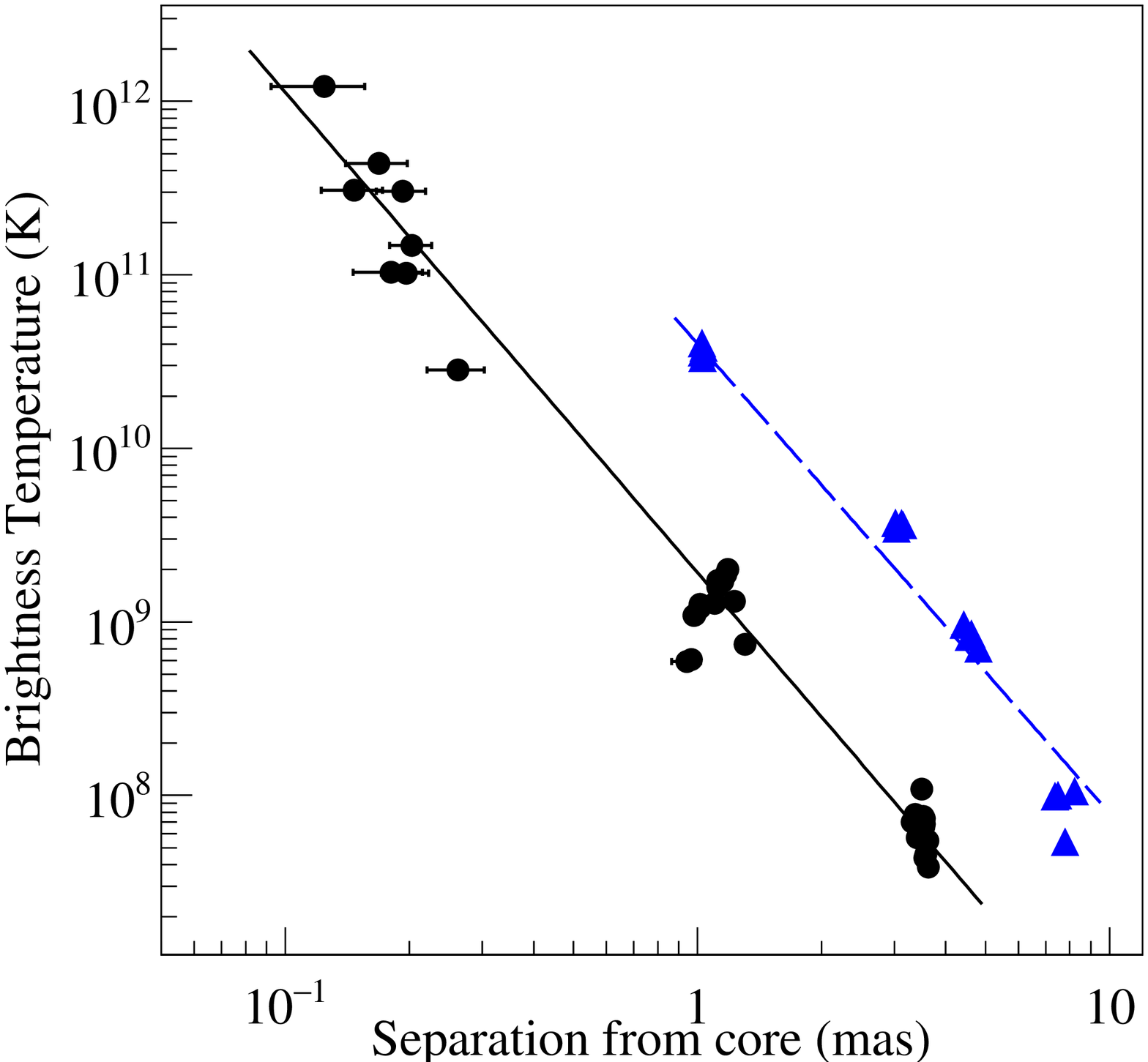}
\includegraphics[scale=.34]{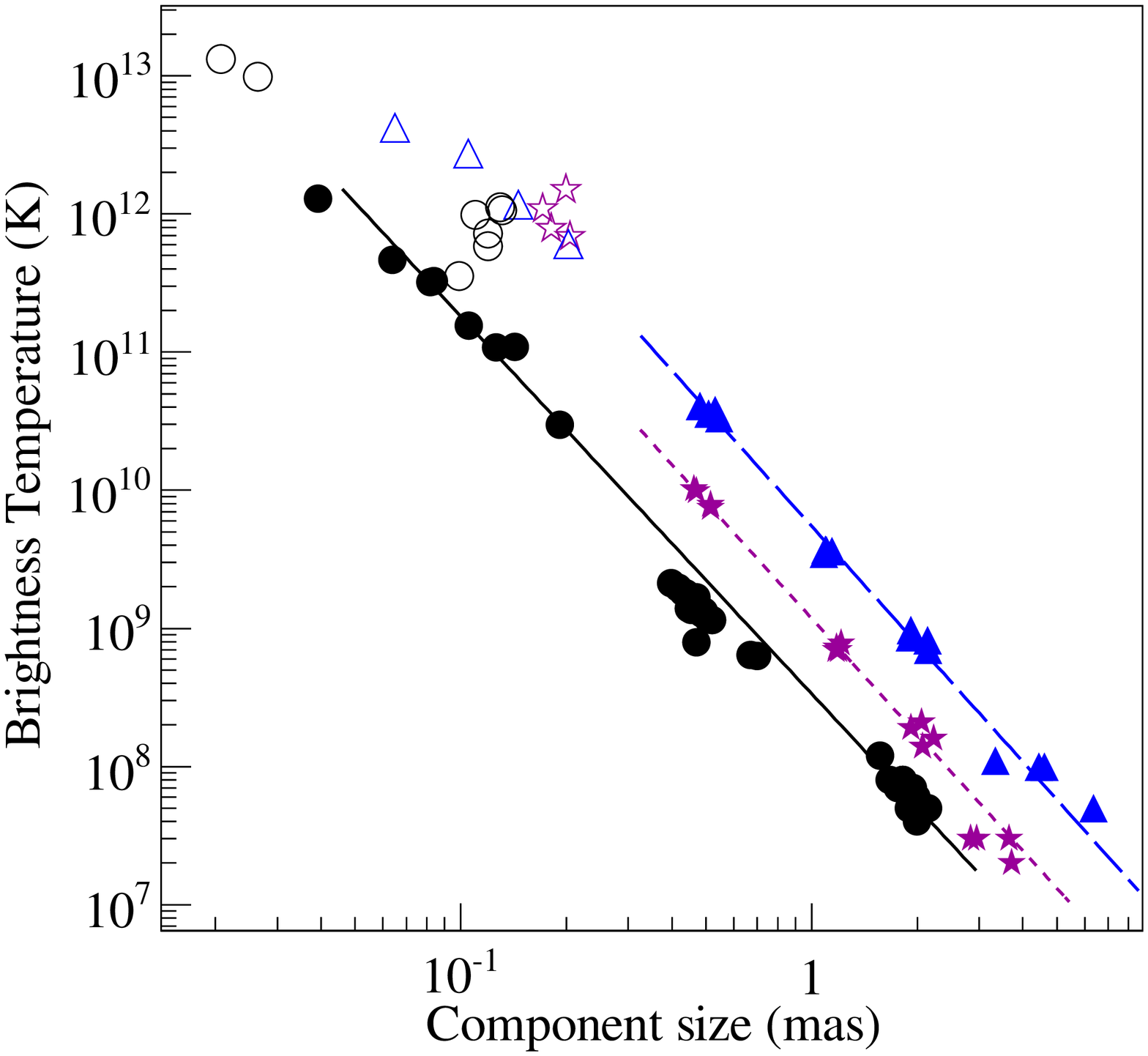}
\caption{(top) Brightness temperature versus distance from the apparent core position at 15.4~GHz (black circles) and 4.6~GHz (blue triangles). 
The power-law fits are shown by lines. Details of the fit are given in Section~\ref{s:tbr}.
(bottom) Brightness temperature versus size of modelled components at 15.4~GHz (black circles), 8.1~GHz (violet stars) and 4.6~GHz (blue triangles). Data at 8.1~GHz is added for comparison. Core components are given by empty symbols and have not been considered in the analysis. The power-law fits are shown by lines. Details of the fit are given in Section~\ref{s:tbr}.}
\label{fig_tbs}
\end{figure}

The brightness temperature~\citep[e.g.,][]{kovalev_etal05} of a slightly resolved component in the source rest frame is given by
\begin{equation}
T_b = 1.22\times10^{12}{\frac{S_\mathrm{comp}(1+z)}{\theta_\mathrm{maj}\theta_\mathrm{min}\nu^2}} \mathrm{K}.
\end{equation}
The $\theta_\mathrm{maj}$ and $\theta_\mathrm{min}$ in the equation are FWHMs of the modelled Gaussians components along the major and minor axes in mas, $S_\mathrm{comp}$ is the flux density of the component in Jy, and the observed frequency $\nu$ is in GHz.
If the component is unresolved, then the upper limit can be used, according to equation~(\ref{eq:limitsize}).
The behavior of the component's brightness temperature with time at 15.4~GHz is shown in Fig.~\ref{fig_tb}, while the evolution of the brightness temperature with the distance from the core ($r$) and with the component's size ($d$) are shown in Fig.~\ref{fig_tbs}.

Under an assumption of a stable conical jet with constant Lorentz factor of the emitting electrons and power-law dependences of the particle density, magnetic field strength, and jet transverse size, the brightness temperature for optically thin synchrotron emission should follow power law  \citep[see ][]{kadler_etal04}
\begin{equation}
T_\mathrm{b,jet} \propto r^{-f},  T_\mathrm{b,jet} \propto d^{-\xi},  f = -l + n + b(1-\alpha),
\label{eq:tb_d}
\end{equation}
where $l$, $n$ and $b$ are power-law indices of the gradient in the jet transverse size ($d \propto r^l$), electron energy distribution ($N_e \propto r^{-n}$), and magnetic field distribution ($B \propto r^{-b}$).
The fit over all epochs results in $f=2.77\pm0.04$ at 15.4~GHz (where 14 epochs are available) and $f=2.71\pm0.02$ at 4.6~GHz (where the extended structure of the source is observed).
The fitted power index $\xi$ equals to 2.73 at 15.4~GHz and 2.83 at 4.6~GHz.
Thus, our data well meet a power-law decrease of the brightness temperature with distance and component's size.
The derived value of the power-law $f$ at parsec scales agrees with theoretical predictions for a straight and a continuous conical jet \citep{kadler_etal04,pushkarev_kovalev_12}, although the brightness temperature of the jet beyond 5~mas downstream experiences a deviation from the power-law, where the jet bends.

\subsection{Core shift analysis}
\label{s:csh}
We studied the apparent frequency-dependent shift of the radio core position \citep{marcaide_shapiro_84,L98}. 
We aligned the images using a method of 2D cross-correlation of the optically thin emission regions in the image plane \citep[e.g.,][]{walker_etal2000,og_09cs}.
The method, which uses a modelled optically thin component as a reference point, did not provide good results, because of (i) ongoing flaring activity, (ii) low flux density in the extended jet component of the source at high frequencies, and (iii) blending of the core with the newly-born component.

\begin{figure}
\centering
\includegraphics[scale=.36]{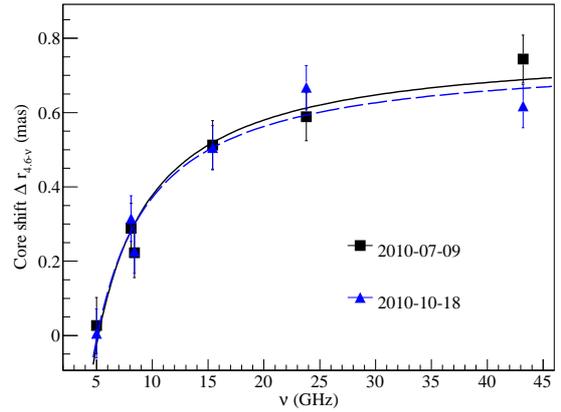}
\caption{Measured value of the core shift relative to 4.6~GHz at two selected epochs. The curves represent a fit to the function $a\nu^{-1}+b$ (solving for $a$ and $b$). The values of the fit are shown in Table~\ref{tb_csh}.}
\label{fig_csh}
\end{figure}

\begin{table*}
\caption{Core shift values in milliarcseconds relative to the 4.6~GHz at four epochs.\label{tb_csh}}
\begin{center}
\begin{tabular}{ccccccc}
\hline
Date & \multicolumn{6}{c}{Frequency (GHz)}\\
\cline{2-7}
& 5.0 & 8.1 & 8.4 & 15.4 & 23.8 & 43.2\\
\hline
2010--05--24&0.02$\pm$0.07&0.14$\pm$0.07&0.16$\pm$0.07&0.38$\pm$0.07&0.61$\pm$0.06&0.64$\pm$0.06\\
2010--07--09&0.03$\pm$0.08&0.29$\pm$0.07&0.22$\pm$0.07&0.51$\pm$0.07&0.59$\pm$0.06&0.74$\pm$0.06\\
2010--08--28&0.03$\pm$0.07&0.26$\pm$0.07&0.30$\pm$0.07&0.39$\pm$0.06&0.70$\pm$0.06&0.63$\pm$0.06\\
2010--10--18&0.01$\pm$0.07&0.31$\pm$0.06&0.23$\pm$0.06&0.51$\pm$0.06&0.67$\pm$0.06&0.62$\pm$0.06\\
\hline
\end{tabular}
\end{center}
\end{table*}

The core shift was calculated as $\Delta\vec{r} = \Delta\vec{r}_{12} - (\vec{r}_1 - \vec{r}_2)$, where $\vec{r}_1$ and $\vec{r}_2$ are positions of the core relative to the phase center of the images at two frequencies, and $\Delta\vec{r}_{12}$ is the displacement between the phase centers of the images at two frequencies. 
The measures of displacement $\Delta\vec{r}_{12}$ are done with the 2D cross-correlation method with the accuracy to be equal the pixel size of the convolved images, 0.05~mas. 

The measured values of the core shift are given in Table~\ref{tb_csh} for all epochs.
The frequency dependence of the core shift is shown in Fig.~\ref{fig_csh} for the 2010--07--09 and 2010--10--18 epochs, when a satisfactory fit of the data can be made.
The shape of the curve can be described by a $r_c(\nu) = a + b{\nu}^{-1/k_r}$ law~\citep{L98}, where $r_c(\nu)$ is position of the core at frequency $\nu$, and coefficient $k_r$ holds information about physical conditions in the ultracompact jet region:
\begin{equation}
k_r = \frac{(3-2\alpha)b+2n-2}{5-2\alpha}.
\label{eq:kr}
\end{equation}
The indices $b$ and $n$ are the same as in equation~(\ref{eq:tb_d}).

We fixed the value of $k_r = 1$, which is applicable if (i) the dominating absorption mechanism is synchrotron self-absorption, (ii) the jet has a conical shape, and (iii) the equipartition the particle and magnetic field densities holds~\citep[see, e.g., ][]{blandford_koenigl_79,L98,hirotani_05,zamaninasab_etal14,zdziarski_etal15}.
Our observations agree well with $k_r = 1$~\citep[e.g., ][]{og_09cs,sokolovsky_etal11,zdziarski_etal15}.
The two-frequency core position offset $\Omega_{r\nu}$ then is given by~\citep{L98}

\begin{equation}
\Omega_{r\nu} = 4.85\cdot10^{-9}{\frac{\Delta r D_\mathrm{L}}{(1+z)^2}\cdot{\frac{\nu_1\nu_2}{\nu_2 - \nu_1}}} \mathrm{~pc\cdot GHz},
\label{eq:cshoffset}
\end{equation}
where $\Delta r$ is measured core shift at two frequencies $\nu_1$ and $\nu_2$ ($\nu_1 \textless \nu_2$, GHz) in mas, and the luminosity distance is measured in parsecs. 
We calculated the average $\Omega_{r\nu}$ value through the 15.4--8.4, 15.4--8.1, 15.4--5.0, and 15.4--4.6~GHz pairs, using the 2010--07--09 and 2010--10--18 epochs, considering $k_r=1$. 
This results in a value of (32$\pm$8)~pc$\cdot$GHz, which is similar to the findings of \citet{pushkarev_etal12} in a large number of sources.

\subsection{Linear polarization}
\label{s:lp}
S4~1030+61 shows linear polarization in the jet and core regions at 4.6 and 5.0~GHz, and only in the core region at other frequencies. 
The evolution of the polarized flux density and degree of polarization with time and frequency is given in Table~\ref{tb_fp} and Table~\ref{tb_15pol}. 
Both the linearly polarized flux density and the degree of linear polarization at 15.4~GHz do not show correlation with the total flux density. 

The degree of polarization shows a peaked shape (see Fig.~\ref{fig_m_l}) with wavelength, with the maximum around 4.5~cm, indicating the complex polarized structure of the source. 
It was shown~\citep[e.g.,][]{farnes_etal14} that a Gaussian function reproduces well the wavelength-dependence of various number of Faraday effects~\citep[][]{sokoloff_etal98}. 
The resulting Gaussian fit, given in  Fig.~\ref{fig_m_l}, is very close to the anomalous depolarization~\citep[][]{sokoloff_etal98,homan_12} or to the smearing of multiple RM components~\citep[][]{conway_etal74}.

\subsection{Faraday rotation}

\begin{figure}
\centering
\includegraphics[scale=.4]{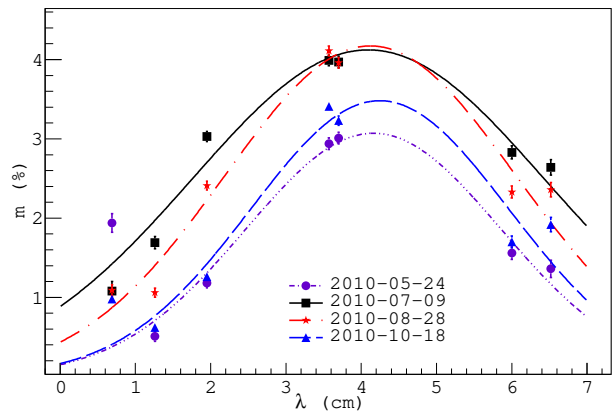}
\caption{Degree of polarization vs. wavelength in the core region at four epochs. Results of a Gaussian fit to the data points are shown. Details are given in Section~\ref{s:lp}.}
\label{fig_m_l}
\end{figure}

In order to restore the intrinsic direction of polarization we calculated the Faraday rotation measure \citep[RM,][]{B66,T91}. 
To perform the Faraday analysis we split the full frequency range on three ranges: 4.6-8.4~GHz, 8.1-15.4~GHz and 15.4-43.2~GHz because of (i) the significant difference in the image resolution, (ii) the complex behavior of degree of polarization with wavelength (see Fig.~\ref{fig_m_l}) and (iii) opacity effects in this region. 
All images were convolved with the appropriate beam size and coaligned using the optically thin components (see Section~\ref{s:csh}).
We assume a linear dependence of electric vector position angle $\phi$ (EVPA) from squared wavelength in frequency intervals:
\begin{equation}
\phi = \phi_0 + RM\cdot\lambda^2,
\end{equation}
where $\phi_0$ is the intrinsic source's EVPA. In such case only lower estimates on the value of Faraday RM can be obtained.

The value of the Faraday rotation of the our Galaxy interstellar medium measured to be of $10.2\pm9.6$~\rmu \citep{taylor_etal09}.
All results given in this paper do not include separate compensation for the Galaxy rotation. 

\begin{figure}
\centering
\includegraphics[angle=-90,scale=.32]{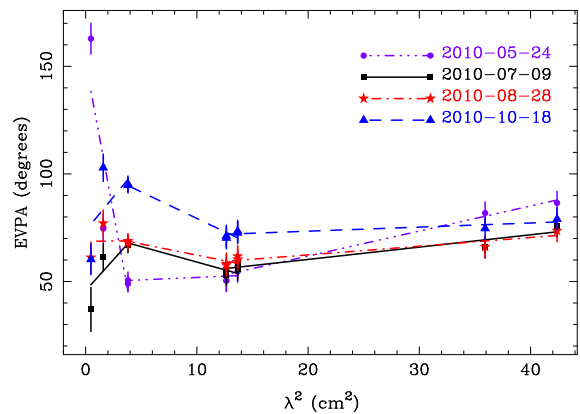}
\caption{Electric vector position angle at the central pixel of the image versus squared wavelength (points) and fitted Faraday rotation (lines) at four epochs. The images at the 4.6--8.4~GHz, 8.1--15.4~GHz, and 15.4--43.2~GHz ranges have been convolved with the beam size and resolution of 4.6, 8.1, and 15.4~GHz maps respectively.
The fitted values to the Faraday rotation are given in Table~\ref{tb_frm}.}
\label{fig_frm}
\end{figure}

\begin{table}
\begin{center}
\caption{Faraday rotation in the core region (\rmu).\label{tb_frm}}
\begin{tabular}{cccc}
\hline
Date & \multicolumn{3}{c}{Frequency ranges (GHz)}\\
\cline{2-4}
& 4.6--8.4 & 8.1--15.4 & 15.4--43.2\\
\hline
2010--05--24 & $203\pm36$ & $40\pm50$ & $-4874\pm152^a$\\
2010--07--09 & $100\pm49$ & $-250\pm73$ & $1004\pm46$\\
2010--08--28 & $71\pm54$ & $-181\pm73$ & $12\pm69$\\
2010--10--18 & $33\pm70$ & $-430\pm104$ & $1093\pm74^a$\\
\hline
\multicolumn{4}{l}{$^a$ The data points have significant deviation from the linear fit}\\
\end{tabular}
\end{center}
\end{table}

The estimated values of the rotation measure are given in Fig.~\ref{fig_frm} and summarized in Table~\ref{tb_frm}.
The assumption about linear dependence is valid for the fit at 1.4--15.4~GHz range, where the rotation may be attributed to the medium, located in front of the emission region (external screen).
Meanwhile, we see significant departure from the linear dependence at higher frequencies (15 to 43 GHz). 
While a least-squares linear regression fits the data satisfactorly in this range at epochs 2010--07--09 and 2010--08--28, a significant deviation from the linear dependence is seen at 2010--05--24 and 2010--10--18, even if n$\pi$-rotations have been applied.
Thus, the provided measured values of the RM in Table~\ref{tb_frm} should be considered as lower limits.

If the anomalous depolarization takes place, then the core of S4~1030+61 posses twisted magnetic field with reversals, which may explain changes in RM sign. 
If this region contains multiple RM components, then the degree of polarization will show an oscillatory behavior with wavelength and the values of RM will differ from the defined ones.

\subsection{Direction of polarization}
\begin{figure*}
\centering
\includegraphics[angle=-90,scale=.7]{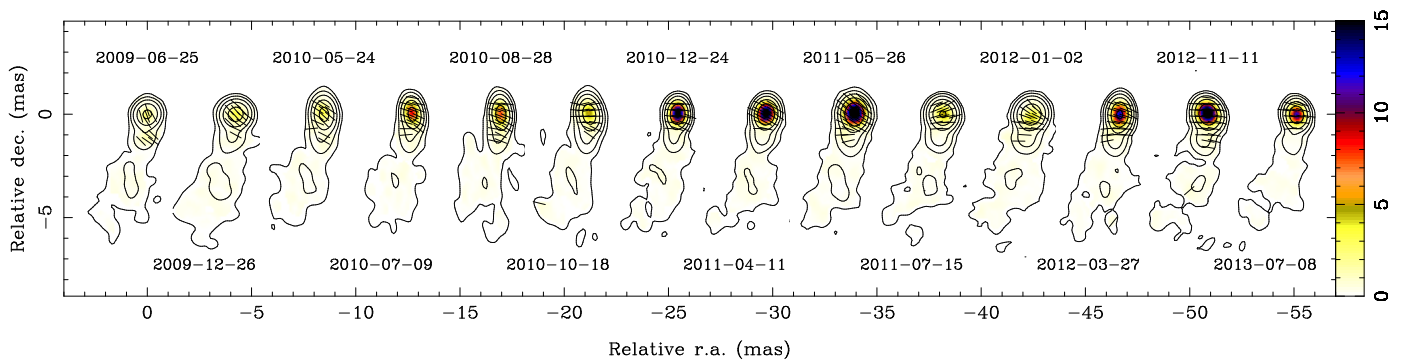}
\caption{S4~1030+61 naturally weighted contour images of total intensity at 15~GHz from 2009 (left) to 2013 (right) for all the MOJAVE epochs. 
Contours of equal intensity are plotted starting from 4 r.m.s.\ level at $\times$5 steps. 
The average EVPA uncertainty is $\pm5\degr$.
The overlaid color images show the linearly polarized intensity in mJy, and ticks represent the direction of EVPA, not corrected for Faraday effects. The details are given in Table~\ref{tb_15pol}.}
\label{fig_mfc_u}
\end{figure*}

\begin{figure}
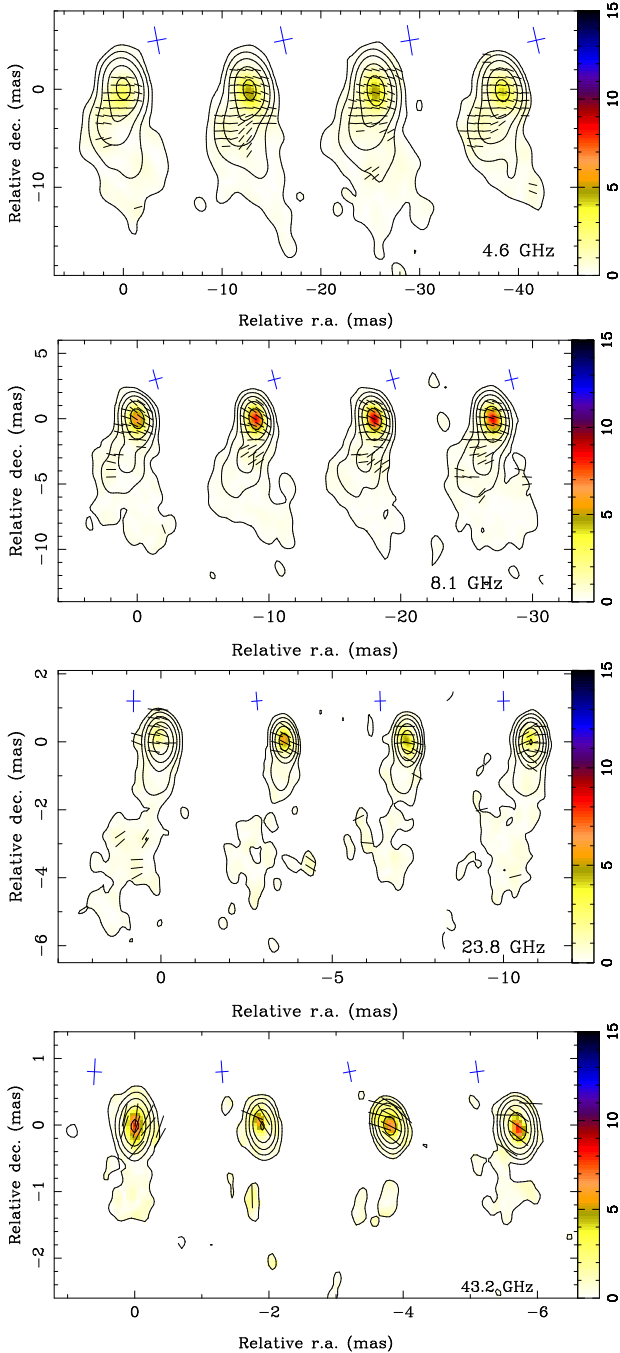

\includegraphics[angle=-90,scale=.32]{mfcore_1030+611_c1_1.eps}\\
\includegraphics[angle=-90,scale=.32]{mfcore_1030+611_x1_1.eps}\\
\includegraphics[angle=-90,scale=.32]{mfcore_1030+611_k1_1.eps}\\
\includegraphics[angle=-90,scale=.32]{mfcore_1030+611_q1_1.eps}
\caption{Observed EVPA orientation (not corrected for RM) at 4.6~GHz, 8.1~GHz, 23.4~GHz and 43.2~GHz at four epochs (from top to bottom). 
The sticks represent the direction of linear polarization. 
The naturally weighted total intensity contours start at 3$\times$rms level and increase by factor of 4. 
The values of rms are given in Table~\ref{tb_fp}. The FWHM of the synthesized beam is shown by crosses.
The overlaid color images show the linearly polarized intensity in mJy and are within the same range as in Fig.~\ref{fig_mfc_u}.}
\label{fig_mfc_all}
\end{figure}

The electric vectors maps, uncorrected for RM, are shown in Fig.~\ref{fig_mfc_u} and Fig.~\ref{fig_mfc_all}.

Faraday effects are less prominent at higher frequencies and we analyse them first. 
The 43.2~GHz EVPA is directed along the jet at 2010--05--24 (see Fig.~\ref{fig_mfc_all}), and rotates on 90\degr and became transverse to the jet direction at later epochs. While EVPA at 23.8~GHz keeps transverse direction through these epochs. This 90\degr jump of EVPA at 43.2~GHz may be due to (i) time variations of the Faraday rotation measure, (ii) smearing of the polarized emissions of emerging component and the core and (iii) change in the opacity at this epoch. If (i) option holds, then RM of 3$\times10^4$rad$\cdot m^2$ is needed to rotate EVPA on 90\degr between 23.8 and 43.2~GHz. If such RM is presented, then it will affect other frequencies, which is not observed.
The behavior of the polarization degree with wavelength (Fig.~\ref{fig_m_l}) is consistent with option (ii), suggesting smearing of the polarized emissions of the core and C3 components through all four epochs. Since the 90\degr jump is seen only during the first epoch, this option is not applicable too.
It is more likely that 90\degr flip is resulted from opacity change as a precursor of radio outburst
\citep[see, e.g.,][]{2010ApJ...710..810A,2011MNRAS.417..359O}.
Since the core of S4~1030+61 is only partially opaque (see Fig. 6), direction of the electric and magnetic fields is coincide. Together with results above this implies transverse magnetic field in the core of the quasar.

Considering these results with 15~GHz MOJAVE EVPA maps (Fig.~\ref{fig_mfc_u}), we see relatively small (tens of degrees) variations of the EVPA during the activity states of the source.
Since these values are uncorrected for Faraday rotation, it can not be distinguished, whether these variations come from RM changes or from changes in orientation of the magnetic field. 
The measured value of the Faraday rotation at 2010--05--24 - 2010--10--18 of few hundreds \rmu\ alters the 15~GHz EVPA on 6.5\degr, and should not change significantly values discussed above.

\begin{table}
\begin{center}
\caption{The 15.4~GHz core polarization degree, linear polarization, its error, and EVPA.\label{tb_15pol}}
\begin{tabular}{ccccc}
\hline
Date & m & p & $\sigma_p$ & $\phi$ \\
 & (\%) & (mJy) & (mJy) & (deg) \\
\hline
2009--06--25 & 0.97 & 2.35  & 0.21 & 20.91\\
2009--12--26 & 1.27 & 3.35  & 0.16 & 41.89\\
2010--05--24 & 1.11 & 2.84  & 0.17 & 48.51\\
2010--07--09 & 2.95 & 7.47  & 0.17 & 67.52\\
2010--08--28 & 2.35 & 6.73  & 0.17 & 68.88\\
2010--10--18 & 1.28 & 4.71  & 0.16 & 96.46\\
2010--12--24 & 3.64 & 16.91 & 0.15 & 80.44\\
2011--04--11 & 3.34 & 18.68 & 0.17 & 59.05\\
2011--05--26 & 4.02 & 21.63 & 0.17 & 53.79\\
2011--07--15 & 0.78 & 4.86  & 0.15 & 105.02\\
2012--01--02 & 0.75 & 3.81  & 0.16 & 62.88\\
2012--03--27 & 2.09 & 11.67 & 0.14 & 73.48\\
2012--11--11 & 4.63 & 19.90 & 0.15 & 83.68\\
2013--07--08 & 2.95 & 10.97 & 0.16 & 80.02\\
\hline
\end{tabular}
\end{center}
\end{table}


\subsection{Decomposition of the radio light curve into flares}

\label{sec_decomp}
The radio light curves can be considered as a sum of the emission from the quiescent jet and the single flares (outbursts) connected with the ejection of a new VLBI component from the radio core \citep{savolainen_etal02}. 
Under such assumption, the radio light curve can be decomposed into individual exponential flares ~\citep{valtaoja_etal99}.
From the fits, the intrinsic jet parameters can be estimated\citep{lvw_99_ii,lv_99_iii,Hovatta_etal09} using the fastest flare: Lorentz factor ($\Gamma_\mathrm{var}$), Doppler boosting factor (${\delta}_\mathrm{var}$), and the viewing angle ($\theta_\mathrm{var}$). 
Following this idea, the radio flux density can be represented as the sum of the flares with the shape given by
\begin{equation}
\Delta S(t) = \left\{
     \begin{array}{lr}
       \Delta S_\mathrm{max}e^{(t-t_\mathrm{max})/\tau},\qquad t < t_\mathrm{max},\\
       \Delta S_\mathrm{max}e^{(t_\mathrm{max}-t)/1.3\tau},\quad t > t_\mathrm{max},
     \end{array}
   \right.
\end{equation}
where $\Delta S_\mathrm{max}$ is the maximum amplitude of the flare in Jy, $t_\mathrm{max}$ is the 
epoch of the flare maximum and $\tau$ is the rise time of the flare in days. 
The lower limit on the observed variability brightness temperature of the source (in its proper frame) for a homogeneous sphere is given by
\begin{equation}
T_{\mathrm{b,var}} = 1.548\times10^{-32}{\frac{\Delta S_\mathrm{max}D^2_L}{\nu^2\tau^2(1+z)}} \mathrm{K},
\end{equation}
where $D_L$ is the luminosity distance in meters, $\nu$ is the observed frequency in GHz, and $z$ is the redshift.

The variability Doppler factor is defined as
\begin{equation}
{\delta}_\mathrm{var} = \Big( {\frac{T_\mathrm{b,var}}{T_\mathrm{b,int}}}\Big)^{1/3},
\end{equation}
where the intrinsic brightness temperature $T_\mathrm{b,int}$ is equipartition temperature,  $10^{11}$~K~\citep{readhead_94}. 
Taking into account apparent superluminal velocities of the jet components $\beta_\mathrm{app}$, the variability Lorentz factor $\Gamma_\mathrm{var}$ and the jet viewing angle $\theta_\mathrm{var}$ can be calculated using following equations:
\begin{equation}
\Gamma_\mathrm{var} = {\frac{\beta^2_\mathrm{app}+{\delta}^2_\mathrm{var}+1}{2{\delta}_\mathrm{var}}},
\end{equation}
\begin{equation}
\theta_\mathrm{var} = \mathrm{arctan}{\frac{2\beta_\mathrm{app}}{\beta^2_\mathrm{app}+{\delta}^2_\mathrm{var}-1}}.
\end{equation}

The result of the decomposition of the 4-years 15~GHz light curve is shown on Fig.~\ref{fig_ovro}. 
Estimate of the median value of $v_\mathrm{var}$ over all flares accounts 9. Meanwhile selection of the fastest flare may give better estimate of Doppler factor, since the flare likely reaches limiting brightness temperature and does not suffer from smearing with other flares \citep[see, e.g.,][]{Hovatta_etal09}.
The estimated fastest flare reaches maximum within 2011--11--17 and 2011--11--19 and corresponds to the maximum amplitude and flare rise timescale of 183$\pm$10 mJy and (39$\pm$7 days). 
This gives $T_\mathrm{b,var} = (3.55\pm1.14)\times10^{14}$~K and an estimate of ${\delta}_\mathrm{var}=15$. This value is considered to be constant along the jet.

S4~1030+61 experiences its largest flares at 2013--09--27 and 2011--03--07.
We do not have enough information to analyze the flare of 2013--09--27, meanwhile our data cover well the 2011--03--07 flare.
The modelling of the 15~GHz data shows that before 2011--03--07 the core has an elliptical shape. 
Meanwhile, after the flare reached a maximum, the model of the source in the core region became more reliable with the introduction of the C3 component. 
\cite{savolainen_etal02} show that time ejection of the new components in AGN jets is close to the peak of a flare. Our modelling shows that C3 was ejected in the middle of 2007, which makes the connection of the 2011--03--07 flare with C3 to be unlikely.
It is supported by Fig.~\ref{fig_kin}, which shows that flaring activity in the radio band after 2011--03--07 takes place in the core.

To calculate jet parameters, we consider the fastest measured velocity in the jet of S4~1030+61, 6.4$c$, and the velocity of a newly born component, 2.7$c$.
It results in $\Gamma_\mathrm{var} = 9.0\pm1.1$ and $\theta_\mathrm{var} = 2.7\degr\pm0.6\degr$ and $\Gamma_\mathrm{var} = 7.9\pm1.0$ and $\theta_\mathrm{var} = 1.3\degr\pm0.4\degr$ for these velocities accordingly, and are typical for \textit{Fermi}-detected quasars~\citep[e.g.,][]{savolainen_etal10,lister_etal11}

\subsection{Connection of the radio and $\gamma$-ray regions}
\label{s:c_g_r}
\begin{figure}
\centering
\includegraphics[scale=.6]{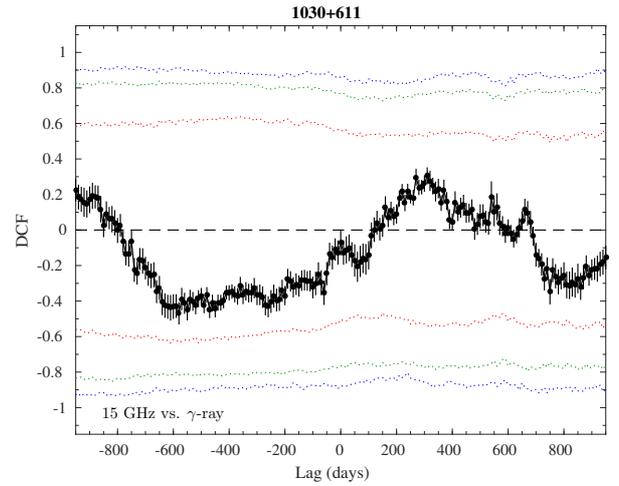}
\caption{DCF obtained from the correlation of radio and weekly binned $\gamma$-ray light curves in the frame of observer for S4~1030$+$61. 
The dotted lines correspond to 1$\sigma$, 2$\sigma$, and 3$\sigma$ significance levels, and are given in red, green and blue with distance from the zero level respectively.}
\label{fig_cdf}
\end{figure}

In S4~1030+61, prominent variability in the $\gamma$-ray is observed during 2010--04--01 and 2011--11-25
and in the middle of the year 2013 (see Fig.~\ref{fig_ovro}). 
To investigate the connection between the radio and $\gamma$-ray variability, we performed a cross-correlation analysis using the discrete correlation function \citep[DCF;][]{edelson_krolik_88}, for the unevenly sampled light curves. 
The DCF is estimated as
\begin{equation}
{\rm DCF}_{ij} = \frac{(a_i-\bar{a_{\tau}})(b_j-\bar{b_{\tau}}) }{\sigma_{a\tau} \sigma_{b\tau}},
\end{equation}
where $a_i, b_j$ are the observed fluxes at times $t_i$ and $t_j$, and $\bar{a_{\tau}}, \bar{b_{\tau}}, \sigma_{a\tau}$ and $\sigma_{b\tau}$ are the means and standard deviations of the points in the overlapping time lag bins, which constrains the DCF within the [$-1,+1$] interval \citep[e.g.,][]{welsh_99}. The final DCF obtained from the average of the bins is shown in Fig.~\ref{fig_cdf}. 
The significance of the DCF peak is then tested by cross-correlating 1000 simulated light curves at radio and $\gamma$-ray bands. 
The light curves are simulated with a power-law slope of 2.75 and 1.70 at radio and $\gamma$ rays, respectively.
From the distribution of the DCF at every time lag bin, we estimated the 68.27 per cent ($1\sigma$), 95.45 per cent ($2\sigma$) and 99.73 per cent ($3\sigma$) significance levels. 
These are denoted by the red, green, and blue lines, respectively, in the Fig.~\ref{fig_cdf}. 
For details of the procedure used in estimating the power-law slope of the light curves and the significance of the cross-correlation, see \citet{ramakrishnan_etal15}.

The correlation between the radio and $\gamma$-ray light curves is insignificant. 
By visual inspection of the light curves (Fig.~\ref{fig_ovro}) it is evident that both bands experience enhanced activity around the same time. 
From Fig.~\ref{fig_kin} we can also see that the radio flaring is due to the flaring of the core component, as was also suggested to be the site of flaring in both radio and $\gamma$-ray bands by \cite{kovalev_etal09} and \cite{pushkarev_etal10}. 
The first and strongest $\gamma$-ray flare peaks at 2010--04--15 and the first large radio flare during this activity period at 2011--03--07. 
If we assume that the  events are connected, we can estimate the spatial separation between the emitting regions at these wavelengths.
Following \cite{pushkarev_etal10} the distance between the locations of $\gamma$-ray and radio core emission is
\begin{equation}
\Delta r = {\frac{{\delta} \Gamma \beta c \Delta t_\mathrm{obs}}{\mathrm{sin}\theta (1+z)}}, 
\end{equation}
where $\Delta t_\mathrm{obs}$ is the observed time delay between two wavelengths.
Considering the time lag between $\gamma$-ray and radio peaks of 0.8 years, it results in a deprojected distance of the $\gamma$-ray emitting region from the 15~GHz radio core of about 12~pc.
The suggested $\gamma$-to-radio delay and the estimated distance are typical for $\gamma$-bright quasars \citep{pushkarev_etal10}.

\section{Discussion}
S4~1030+61 shows flaring activity at radio wavelengths through the five years of observations without reaching a quiescent state. 
Meanwhile at $\gamma$ rays it has two prominent activity periods.
The OVRO light curve (Fig.~\ref{fig_ovro}) shows that the source experienced its largest flare in early 2014, and may be connected with enhanced activity in the $\gamma$-ray band.
Despite the strong activity of the source, our kinematic analysis at 15~GHz does not provide evidence for the birth of another component.
The C1 component seems to be stationary, that can be explained by (i) changes of the angle between the line of the jet propagation and the line of the sight (bending of the jet), (ii) interaction of the component with the boundary between the jet outflow and the interstellar medium \citep[e.g.,][]{homan_etal03} or by (iii) a standing shock in a jet. 
The decrease of the brightness temperature with distance at 15~GHz well follows a power-law up to the location of the C1 component (see Fig.~\ref{fig_tbs}), which places the region where jet bends, further downstream the jet.
This makes option (i) to be unlikely.
The spectral index maps show optically-thin emission at the location of C1 (see Fig.~\ref{fig_spi}), which also rules out option (ii).

The observed value of the brightness temperature ($T_\mathrm{obs}$) may be connected to the intrinsic temperature ($T_\mathrm{int}$) via the Doppler factor: $T_\mathrm{obs} = {\delta}T_\mathrm{int}$. 
Taking into account the estimate of ${\delta}=15$, the average core brightness temperature is close to the equipartition value of $10^{11}$~K during the first six VLBA epochs (2009--06--25 - 2010--10--18), where the source shows a moderate flux variations.
Meanwhile, the average brightness temperature after the separation of C3 components is $3.3\times10^{11}$~K.
\cite{homan_etal06} show that the AGN jet cores are close to equipartition in their median-low state, meanwhile jets at their maximum brightness state go out of equipartition. 
The upper limit of $2\times10^{11}$K in intrinsic brightness temperature reported by the authors is close to our estimate.
They suggest that jets become particle dominated during the flaring activity, resulting from particle injection or acceleration at the jet base.
Hardening of the S4~1030$+$61 spectrum during the outburst evolution at the end of 2010 (see Fig.~\ref{fig_spi}) supports this suggestion.

The brightness temperature of the source's jet decreases with the distance $r$ from the jet base and the size of the components $d$ as $T_\mathrm{b,jet}\propto r^{-f}$ and $T_\mathrm{b,jet}\propto d^{-\xi}$ accordingly. The estimated power-law indices at 4.6 and 15.4~GHz are $f=2.75\pm0.04$ and $\xi=2.8$.
Considering these results, the power-law index of the jet width decrease with distance from the core $l$ can be estimated as $f/\xi$ \cite[see][]{pushkarev_kovalev_12} and resulted in the value of $l$ close to 1.
Applying this to equation~(\ref{eq:tb_d}) and (\ref{eq:kr}), together with $\alpha=-0.82\pm0.2$ at the position of C2 component, leads to the values of power-law indices $n=1.7$, $b=1.1$. These results implies that the equipartition between magnetic and particle energies holds in the S4~1030+61 jet, while transverse magnetic field component dominates in the magnetic field density.

The deprojected distance of the apparent radio core from the central black hole can be estimated as
\begin{equation}
r_\mathrm{core} = {\frac{\Omega_{r\nu}}{\nu\mathrm{sin}\theta}} \approx {\frac{\Omega_{r\nu}\sqrt{1+\beta^2_\mathrm{app}}}{\nu}},
\label{eq:csh_abs}
\end{equation}
where $\nu$ is the observed frequency in GHz, and $\Omega_{r\nu}$ is the core position offset (pc$\cdot$GHz), given by equation~(\ref{eq:cshoffset})~\citep{L98}. The resulted value is of (32$\pm$8)~pc$\cdot$GHz.
Taking into consideration the maximum observed apparent speed, the absolute distance of the 15.4~GHz core from the AGN central engine then equals to (14$\pm$3)~pc, which is close to the median value of 13.2~pc measured by~\cite{pushkarev_etal12} for quasars.
Thus, the 15~GHz core lies outside of the broad line region.

Estimation of the magnetic field strength at 1~pc can be made through the proportionality \citep{hirotani_05,og_09mf,zdziarski_etal15}
\begin{equation}
B_1 \simeq 0.025 \Big( \frac{\Omega^3_{r\nu}(1+z)^3}{{\delta}^2\phi \mathrm{sin}^2\theta} \Big)^{1/4} G,
\end{equation}
where $\phi$ is the half-openning angle of the jet. Considering $2\phi \simeq 0.26\Gamma^{-1}$ \citep{2009AA...507L..33P}, the magnetic field strength estimate at 1~pc is of 2.2~G. The magnetic field strength in the core at observed frequency can be found as 
\begin{equation}
B_\mathrm{core}(\nu) = B_1 r^{-1}_\mathrm{core}
\end{equation}
and for 15.4~GHz core results in 0.16~G.

If we assume that the 2010--04--15 $\gamma$-ray flare and the 2011--03--07 radio flare are connected, then the time lag of 0.8~years between these flares (see Section~\ref{s:c_g_r}) resulted in an estimate of deprojected distance of the $\gamma$-ray emitting region from the 15~GHz core of about 12~pc. 
The core shift measure (see above) places the 15~GHz core at absolute distance of (14$\pm$2)~pc from the central engine. Applying these results, we interpret that the $\gamma$-ray emission originates at 2~pc from the central engine.

The cross-correlation of the radio and $\gamma$-ray light curves is not significant. 
This could be due to the much faster variability time scales at the $\gamma$-ray band. 
Inspecting Fig.~\ref{fig_ovro} by eye, it is evident that both bands experience enhanced activity around the same time. 
The lack of detection of a VLBI component at this period could be due to dissipation of the disturbance, causing $\gamma$-ray emission, on the way from $\gamma$-ray production region to the radio core position.
The ongoing $\gamma$-ray activity after the strongest flare may be due to scattering of the seed photons on the jet medium.
~\cite{marscher_etal10a,marscher_etal10b} suggest that this medium might be a relatively slow sheath of the jet or a faster spine from observations of PKS~1510--089. Meanwhile, the $\gamma$ rays from the main flare are caused by inverse Compton scattering.

The short variability interval of the $\gamma$-ray flux suggests compactness of the emitting region. 
Following \cite{jorstad_etal05}, the size of the emitting region ($R$) cannot be larger than $R < \Delta t \delta c$, where $\Delta t$ is the minimum variability timescale.
The minimum detectable variability, seen on the $\gamma$-ray light curve is within 2 weeks (and limited by \Fermi\ detectability), which implies $R$ to be less than 0.18~pc.
The size of the 43.2~GHz radio core is $\leq0.12$~pc (considering the size of the modelled core to be 0.014~mas). 
The deprojected distance of the 43.2~GHz core to the central engine, following  equation~(\ref{eq:cshoffset}) and equation~(\ref{eq:csh_abs}), is estimated to be 1.3~pc.
This value is close to the estimated earlier possible location of the $\gamma$-ray emission region of 2~pc from the central engine.
It implies that the $\gamma$-ray emission may be localized in the 7~mm core region.

Variations of the parsec-scale jet orientation with time and distance from the jet base have been observed in many blazars \citep[e.g.,][]{stirling_etal03,agudo_09,lister_etal13}.
\citet{rani_etal14} relate these variations with the $\gamma$-ray flux variability, and received significant correlation between them.
Structural changes in jet orientation of S4~1030+61 are seen in Fig.~\ref{fig_cxust}.
Its PA changes from about $\thicksim$166\degr\ (at 5 mas from the core) to 170\degr\ (at the position of C2) and to about $-170$\degr\ (at the position of C3).
Such variations might cause an absence of a correlation between radio and $\gamma$-ray bands in S4~1030+61: changes of jet orientation on sub-parsec-scales after strongest $\gamma$-ray flare may cause $\gamma$-ray flux to be too faint for its detection by \Fermi.

The core region of the source is strongly affected by Faraday effects, which result in the complex behavior of the linear polarization degree with $\lambda$.
The dependence differs from the expected behavior for the optically thick region of a jet \citep{PS67}.
Possible explanations for such behavior are the following: 
(i) anomalous ~\citep{sokoloff_etal98} or inverse~\citep{homan_12} depolarization, which appears in a regular twisted or tangled magnetic fields;
(ii) spectral depolarization~\citep{conway_etal74} due to smearing of multiple components within the observed region.
Unfortunately, distinguishing between these alternatives is difficult. The EVPA vs. $\lambda^2$ behavior is consistent with both hypotheses. 
If the option (i) holds, then the relatively similar behavior of the degree of polarization in time implies a constant field pitch angle during the radio flare. 
The change in the amplitude then may imply a change in the strength of the magnetic field during the flare, which is supported by observed changes in the Faraday rotation.
In turn, the RM is connected with both the magnetic field along the line of sight, and the electron density.
Though \cite{BM10} pointed out that the RM in the core region should be treated cautiously. 
Quantities there change on scales much smaller than the observed interferometric beam, thus all characteristics will be smeared.
The new component will undergo compression while passing the radio core resulting in changes both in strength of magnetic field and electron density.
~\citet[and references therein]{dammando_etal13} show that the source at any active state could be described by changing only in the electron distribution.
The option (ii) can be true, since the core shows inverted spectrum and the C3 component is too close to the core to be studied separately.

\citet{taylor_00} and \citet{zt_01} also observed temporal variations in polarized intensity and RM value in the cores of 3C~273 and 3C~279. 
The authors relate these changes to the creation and ejection of new components there.
Indeed, \cite{lister_etal13} report on emergence of the new components in these sources at times close to  observing epochs of \cite{taylor_00} and \cite{zt_01}.
\citet{lico_etal14} and \citet{giroletti_etal15} show significant temporal variations of polarized flux density, RM and direction of intrinsic EVPA in the core of Markarian~421 during its $\gamma$-ray activity, which connects the magnetic field and $\gamma$-ray emission.
Assuming RM changes of 400~\rmu (twice the measured value in 8.1--15.4~GHz range) in the core of S4~1030$+$61 through the 14 observational epochs, the 15.4~GHz EVPA then should rotate on about 9\degr, which has no significant influence on the $\phi_0$ relative to the observed EVPA, given in Fig.~\ref{fig_uall}.
This likely connects changes in EVPA at 15.4~GHz with the orientation of the magnetic field. 
No clear connection of the magnetic field direction with the activity state can be made.

\section{Summary}

We conducted a radio and $\gamma$-ray study of the quasar S4~1030+61, using multiwavelength data in the period of 2009--2014. The conclusions are following.

(i) The kinematic analysis shows the presence of three jet components, C3, located within few pc from the 15.4~GHz core moving with an apparent velocity of (2.7$\pm$0.7)c, C2 with (6.4$\pm$0.4)c at distances of 7 to 10~pc from the core, and a stationary component C1 with the estimated velocity is (2$\pm$2)c and located few tens of parsec from the core.
All components were ejected before the time of our observations.

(ii) The decomposition of the 15.4~GHz OVRO radio light curve reveals nine prominent radio flares, with the fastest one occurred at 2011--11--18.
The estimated variability parameters of the jet are as follows: Doppler factor ${\delta}_\mathrm{var}\approx15$, Lorentz factor $\Gamma_\mathrm{var}=9.02\pm1.10$, and viewing angle  $\theta_\mathrm{var}=2.7\degr\pm0.6\degr$.

(iii) The core shift results do not change significantly within three months of our multi-frequency VLBI observations. 
The deprojected distance of the 15.4~GHz core to the central black hole is estimated as 14$\pm$3~pc, and is close to the median value of 13.2~pc measured by~\cite{pushkarev_etal12} for MOJAVE quasars.

(iv) The estimated magnetic field strength at 1~pc from the central engine of the quasar is of 2.2~G and decreases with the distance as $B \propto r^{-1.1}$, while electron density distribution and the gradient in the jet transverse size decrease as $N \propto r^{-1.7}$ and $d \propto r$ respectively.

(v) If the strongest $\gamma$-ray flare is associated with one of the strongest radio flares during 2009--2014, the deprojected distance between the $\gamma$-ray emitting region and the 15.4~GHz apparent core position is estimated to be about 12~pc. 
Together with the core shift analyses it results in the location of the region where $\gamma$-ray emission originates to lie within 2 parsecs from the central black hole.
The shortest $\gamma$-ray variability time-scale may indicate that the $\gamma$-ray emission originates close to the mm radio core.

(vi) We detected changes in the value and direction of linear polarization during the source flaring activity. 
The EVPA shows slight variations of its direction during the flares, with no clear correlation with the activity state or total flux density. 90\degr EVPA flip is observed during the beginning of the strongest radio flare, which is likely resulted from opacity change.
The Faraday rotation measure shows variations in its value with time.
Therefore, changes in the magnetic field strength, the electron density or the size of the region are taking place.

(vii) The value of the intrinsic brightness temperature in the core is estimated to be $>3.3\times10^{11}$~K, which is above the equipartition value.
The brightness temperature gradient with the distance from the core is well approximated by a power-law.  

(viii) Temporal hardening of the core spectrum, the high value of the core brightness temperature and changes in the value of Faraday RM imply that the radio flaring activity of S4~1030+61 is accompanied by injection of relativistic particles and energy losses at the jet base.

Summarizing, our results show that the jet of S4~1030+61 can be well described by standard conical jet theories~\citep[e.g.,][]{blandford_koenigl_79}.

\section*{Acknowledgments}
We thank A.~Pushkarev for useful consultations about a wide range of questions, A.~Kutkin for helpful discussion about radio-to-$\gamma$-ray correlation, E.~Ros for comprehensive comments and corrections as internal reviewer at the Max-Planck-Institut f\"ur Radioastronomie, and the anonymous referee for a thorough reading of the manuscript and valuable comments which helped to improve the paper.
The polarization study was supported by the Russian Foundation for Basic Research (project 14-02-31789). The core shift study was supported by Russian Science Foundation (project 16-12-10481).
T. Hovatta was supported by the Academy of Finland project number 267324.
The luminosity distance of the source was computed using the cosmological calculator \citep{wright_06}.
The Very Long Baseline Array is an instrument of the National Radio Astronomy Observatory, a facility of the National
Science Foundation operated under cooperative agreement by Associated Universities, Inc. 
The OVRO 40-m monitoring program is supported in part by NASA grants NNX08AW31G and NNX11A043G, and NSF grants AST-0808050 and AST-1109911.
This research has made use of data from the University of Michigan Radio Astronomy Observatory (absolute electric vector position angle calibration) which has been supported by the University of Michigan and by a series of grants from the National Science Foundation, most recently AST-0607523.
This research has made use of data from the MOJAVE database that is maintained by the MOJAVE team \citep{lister_etal09}.


\bibliographystyle{mnras}
\bibliography{1030}

\begin{thebibliography}{}
\makeatletter
\relax
\def\mn@urlcharsother{\let\do\@makeother \do\$\do\&\do\#\do\^\do\_\do\%\do\~}
\def\mn@doi{\begingroup\mn@urlcharsother \@ifnextchar [ {\mn@doi@}
  {\mn@doi@[]}}
\def\mn@doi@[#1]#2{\def\@tempa{#1}\ifx\@tempa\@empty \href
  {http://dx.doi.org/#2} {doi:#2}\else \href {http://dx.doi.org/#2} {#1}\fi
  \endgroup}
\def\mn@eprint#1#2{\mn@eprint@#1:#2::\@nil}
\def\mn@eprint@arXiv#1{\href {http://arxiv.org/abs/#1} {{\tt arXiv:#1}}}
\def\mn@eprint@dblp#1{\href {http://dblp.uni-trier.de/rec/bibtex/#1.xml}
  {dblp:#1}}
\def\mn@eprint@#1:#2:#3:#4\@nil{\def\@tempa {#1}\def\@tempb {#2}\def\@tempc
  {#3}\ifx \@tempc \@empty \let \@tempc \@tempb \let \@tempb \@tempa \fi \ifx
  \@tempb \@empty \def\@tempb {arXiv}\fi \@ifundefined
  {mn@eprint@\@tempb}{\@tempb:\@tempc}{\expandafter \expandafter \csname
  mn@eprint@\@tempb\endcsname \expandafter{\@tempc}}}

\bibitem[\protect\citeauthoryear{{Abdo} et~al.,}{{Abdo}
  et~al.}{2010}]{2010ApJ...710..810A}
{Abdo} A.~A.,  et~al., 2010, \mn@doi [\apj] {10.1088/0004-637X/710/1/810},
  \href {http://adsabs.harvard.edu/abs/2010ApJ...710..810A} {710, 810}

\bibitem[\protect\citeauthoryear{{Abdo} et~al.,}{{Abdo}
  et~al.}{2011}]{abdo_etal11}
{Abdo} A.~A.,  et~al., 2011, \mn@doi [\apj] {10.1088/0004-637X/730/2/101},
  \href {http://esoads.eso.org/abs/2011ApJ...730..101A} {730, 101}

\bibitem[\protect\citeauthoryear{{Acero} et~al.,}{{Acero}
  et~al.}{2015}]{acero_etal15}
{Acero} F.,  et~al., 2015, \mn@doi [\apjs] {10.1088/0067-0049/218/2/23}, \href
  {http://adsabs.harvard.edu/abs/2015ApJS..218...23A} {218, 23}

\bibitem[\protect\citeauthoryear{{Agudo}}{{Agudo}}{2009}]{agudo_09}
{Agudo} I.,  2009, in {Hagiwara} Y.,  {Fomalont} E.,  {Tsuboi} M.,   {Yasuhiro}
  M.,  eds,  Astronomical Society of the Pacific Conference Series Vol. 402,
  Approaching Micro-Arcsecond Resolution with VSOP-2: Astrophysics and
  Technologies. p.~330 (\mn@eprint {arXiv} {0803.3388})

\bibitem[\protect\citeauthoryear{{Agudo} et~al.,}{{Agudo}
  et~al.}{2011}]{agudo_etal11}
{Agudo} I.,  et~al., 2011, \mn@doi [\apjl] {10.1088/2041-8205/726/1/L13}, \href
  {http://adsabs.harvard.edu/abs/2011ApJ...726L..13A} {726, L13}

\bibitem[\protect\citeauthoryear{{Blandford} \& {K{\"o}nigl}}{{Blandford} \&
  {K{\"o}nigl}}{1979}]{blandford_koenigl_79}
{Blandford} R.~D.,  {K{\"o}nigl} A.,  1979, \mn@doi [\apj] {10.1086/157262},
  \href {http://adsabs.harvard.edu/abs/1979ApJ...232...34B} {232, 34}

\bibitem[\protect\citeauthoryear{{Blandford} \& {Levinson}}{{Blandford} \&
  {Levinson}}{1995}]{bl_95}
{Blandford} R.~D.,  {Levinson} A.,  1995, \mn@doi [\apj] {10.1086/175338},
  \href {http://adsabs.harvard.edu/abs/1995ApJ...441...79B} {441, 79}

\bibitem[\protect\citeauthoryear{{Broderick} \& {McKinney}}{{Broderick} \&
  {McKinney}}{2010}]{BM10}
{Broderick} A.~E.,  {McKinney} J.~C.,  2010, \mn@doi [\apj]
  {10.1088/0004-637X/725/1/750}, \href
  {http://esoads.eso.org/abs/2010ApJ...725..750B} {725, 750}

\bibitem[\protect\citeauthoryear{{Brun} \& {Rademakers}}{{Brun} \&
  {Rademakers}}{1997}]{root}
{Brun} R.,  {Rademakers} F.,  1997, \mn@doi [Nucl. Instr. Meth. Phys. Res.]
  {10.1016/S0168-9002(97)00048-X}, 389, 81

\bibitem[\protect\citeauthoryear{{Burn}}{{Burn}}{1966}]{B66}
{Burn} B.~J.,  1966, \mnras, \href
  {http://esoads.eso.org/abs/1966MNRAS.133...67B} {133, 67}

\bibitem[\protect\citeauthoryear{{Carrasco}, {Carrami{\~n}ana}, {Escobedo},
  {Recillas}, {Porras}, {Valdes}  \& {Mayya}}{{Carrasco}
  et~al.}{2010}]{carrasco_atel10}
{Carrasco} L.,  {Carrami{\~n}ana} A.,  {Escobedo} G.,  {Recillas} E.,  {Porras}
  A.,  {Valdes} J.~R.,   {Mayya} Y.~D.,  2010, The Astronomer's Telegram, \href
  {http://adsabs.harvard.edu/abs/2010ATel.3062....1C} {3062}

\bibitem[\protect\citeauthoryear{{Ciprini}}{{Ciprini}}{2010}]{ciprini_atel10}
{Ciprini} S.,  2010, The Astronomer's Telegram, \href
  {http://adsabs.harvard.edu/abs/2010ATel.2622....1C} {2622}

\bibitem[\protect\citeauthoryear{{Conway}, {Haves}, {Kronberg}, {Stannard},
  {Vallee}  \& {Wardle}}{{Conway} et~al.}{1974}]{conway_etal74}
{Conway} R.~G.,  {Haves} P.,  {Kronberg} P.~P.,  {Stannard} D.,  {Vallee}
  J.~P.,   {Wardle} J.~F.~C.,  1974, \mnras, \href
  {http://adsabs.harvard.edu/abs/1974MNRAS.168..137C} {168, 137}

\bibitem[\protect\citeauthoryear{{D'Ammando} et~al.,}{{D'Ammando}
  et~al.}{2013}]{dammando_etal13}
{D'Ammando} F.,  et~al., 2013, \mn@doi [\mnras] {10.1093/mnras/stt344}, \href
  {http://adsabs.harvard.edu/abs/2013MNRAS.431.2481D} {431, 2481}

\bibitem[\protect\citeauthoryear{{Daly} \& {Marscher}}{{Daly} \&
  {Marscher}}{1988}]{1988ApJ...334..539D}
{Daly} R.~A.,  {Marscher} A.~P.,  1988, \mn@doi [\apj] {10.1086/166858}, \href
  {http://adsabs.harvard.edu/abs/1988ApJ...334..539D} {334, 539}

\bibitem[\protect\citeauthoryear{{Dotson}, {Georganopoulos}, {Kazanas}  \&
  {Perlman}}{{Dotson} et~al.}{2012}]{dotson_etal2012}
{Dotson} A.,  {Georganopoulos} M.,  {Kazanas} D.,   {Perlman} E.~S.,  2012,
  \mn@doi [\apjl] {10.1088/2041-8205/758/1/L15}, \href
  {http://adsabs.harvard.edu/abs/2012ApJ...758L..15D} {758, L15}

\bibitem[\protect\citeauthoryear{{Edelson} \& {Krolik}}{{Edelson} \&
  {Krolik}}{1988}]{edelson_krolik_88}
{Edelson} R.~A.,  {Krolik} J.~H.,  1988, \mn@doi [\apj] {10.1086/166773}, \href
  {http://adsabs.harvard.edu/abs/1988ApJ...333..646E} {333, 646}

\bibitem[\protect\citeauthoryear{{Farnes}, {Gaensler}  \& {Carretti}}{{Farnes}
  et~al.}{2014}]{farnes_etal14}
{Farnes} J.~S.,  {Gaensler} B.~M.,   {Carretti} E.,  2014, \mn@doi [\apjs]
  {10.1088/0067-0049/212/1/15}, \href
  {http://adsabs.harvard.edu/abs/2014ApJS..212...15F} {212, 15}

\bibitem[\protect\citeauthoryear{{Fomalont}}{{Fomalont}}{1999}]{fomalont_99}
{Fomalont} E.~B.,  1999, in {Taylor} G.~B.,  {Carilli} C.~L.,   {Perley} R.~A.,
   eds,  Astronomical Society of the Pacific Conference Series Vol. 180,
  Synthesis Imaging in Radio Astronomy II. p.~301

\bibitem[\protect\citeauthoryear{{Fromm}, {Ros}, {Perucho}, {Savolainen},
  {Mimica}, {Kadler}, {Lobanov}  \& {Zensus}}{{Fromm}
  et~al.}{2013}]{fromm_etal13}
{Fromm} C.~M.,  {Ros} E.,  {Perucho} M.,  {Savolainen} T.,  {Mimica} P.,
  {Kadler} M.,  {Lobanov} A.~P.,   {Zensus} J.~A.,  2013, \mn@doi [\aap]
  {10.1051/0004-6361/201321784}, \href
  {http://adsabs.harvard.edu/abs/2013A%26A...557A.105F} {557, A105}

\bibitem[\protect\citeauthoryear{{Fuhrmann} et~al.,}{{Fuhrmann}
  et~al.}{2014}]{fuhrmann_etal14}
{Fuhrmann} L.,  et~al., 2014, \mn@doi [\mnras] {10.1093/mnras/stu540}, \href
  {http://adsabs.harvard.edu/abs/2014MNRAS.441.1899F} {441, 1899}

\bibitem[\protect\citeauthoryear{{Giroletti} et~al.,}{{Giroletti}
  et~al.}{2015}]{giroletti_etal15}
{Giroletti} M.,  et~al., 2015, 2014 Fermi Symposium proceedings, \href
  {http://adsabs.harvard.edu/abs/2015arXiv150304597G} {eConf C141020.1}

\bibitem[\protect\citeauthoryear{{G{\'o}mez}, {Marscher}, {Alberdi}, {Jorstad}
  \& {Agudo}}{{G{\'o}mez} et~al.}{2002}]{gomez_etal02}
{G{\'o}mez} J.~L.,  {Marscher} A.~P.,  {Alberdi} A.,  {Jorstad} S.~G.,
  {Agudo} I.,  2002, VLBA Scientific Memo, 30

\bibitem[\protect\citeauthoryear{{Greisen}}{{Greisen}}{2003}]{aips}
{Greisen} E.~W.,  2003, in {Heck} A.,  ed., Astrophysics and Space Science
  Library 285, Information Handling in Astronomy -- Historical Vistas.
  Dordrecht: Kluwer, p.~109

\bibitem[\protect\citeauthoryear{{Hirotani}}{{Hirotani}}{2005}]{hirotani_05}
{Hirotani} K.,  2005, \mn@doi [\apj] {10.1086/426497}, \href
  {http://esoads.eso.org/abs/2005ApJ...619...73H} {619, 73}

\bibitem[\protect\citeauthoryear{{Homan}}{{Homan}}{2012}]{homan_12}
{Homan} D.~C.,  2012, \mn@doi [\apjl] {10.1088/2041-8205/747/2/L24}, \href
  {http://adsabs.harvard.edu/abs/2012ApJ...747L..24H} {747, L24}

\bibitem[\protect\citeauthoryear{{Homan}, {Lister}, {Kellermann}, {Cohen},
  {Ros}, {Zensus}, {Kadler}  \& {Vermeulen}}{{Homan}
  et~al.}{2003}]{homan_etal03}
{Homan} D.~C.,  {Lister} M.~L.,  {Kellermann} K.~I.,  {Cohen} M.~H.,  {Ros} E.,
   {Zensus} J.~A.,  {Kadler} M.,   {Vermeulen} R.~C.,  2003, \mn@doi [\apjl]
  {10.1086/375726}, \href {http://adsabs.harvard.edu/abs/2003ApJ...589L...9H}
  {589, L9}

\bibitem[\protect\citeauthoryear{{Homan} et~al.,}{{Homan}
  et~al.}{2006}]{homan_etal06}
{Homan} D.~C.,  et~al., 2006, \mn@doi [\apjl] {10.1086/504715}, \href
  {http://adsabs.harvard.edu/abs/2006ApJ...642L.115H} {642, L115}

\bibitem[\protect\citeauthoryear{{Hovatta}, {Valtaoja}, {Tornikoski}  \&
  {L{\"a}hteenm{\"a}ki}}{{Hovatta} et~al.}{2009}]{Hovatta_etal09}
{Hovatta} T.,  {Valtaoja} E.,  {Tornikoski} M.,   {L{\"a}hteenm{\"a}ki} A.,
  2009, \mn@doi [\aap] {10.1051/0004-6361:200811150}, \href
  {http://esoads.eso.org/abs/2009A%26A...494..527H} {494, 527}

\bibitem[\protect\citeauthoryear{{Hovatta}, {Lister}, {Aller}, {Aller},
  {Homan}, {Kovalev}, {Pushkarev}  \& {Savolainen}}{{Hovatta}
  et~al.}{2012}]{hovatta_etal12}
{Hovatta} T.,  {Lister} M.~L.,  {Aller} M.~F.,  {Aller} H.~D.,  {Homan} D.~C.,
  {Kovalev} Y.~Y.,  {Pushkarev} A.~B.,   {Savolainen} T.,  2012, \mn@doi [\aj]
  {10.1088/0004-6256/144/4/105}, \href
  {http://adsabs.harvard.edu/abs/2012AJ....144..105H} {144, 105}

\bibitem[\protect\citeauthoryear{{Inada} et~al.,}{{Inada}
  et~al.}{2012}]{2012AJ....143..119I}
{Inada} N.,  et~al., 2012, \mn@doi [\aj] {10.1088/0004-6256/143/5/119}, \href
  {http://adsabs.harvard.edu/abs/2012AJ....143..119I} {143, 119}

\bibitem[\protect\citeauthoryear{{James} \& {Roos}}{{James} \&
  {Roos}}{1975}]{minuit}
{James} F.,  {Roos} M.,  1975, \mn@doi [Comput.Phys.Commun.]
  {10.1016/0010-4655(75)90039-9}, 10, 343

\bibitem[\protect\citeauthoryear{{Jorstad}, {Marscher}, {Mattox}, {Aller},
  {Aller}, {Wehrle}  \& {Bloom}}{{Jorstad} et~al.}{2001}]{jorstad_etal01}
{Jorstad} S.~G.,  {Marscher} A.~P.,  {Mattox} J.~R.,  {Aller} M.~F.,  {Aller}
  H.~D.,  {Wehrle} A.~E.,   {Bloom} S.~D.,  2001, \mn@doi [\apj]
  {10.1086/321605}, \href {http://adsabs.harvard.edu/abs/2001ApJ...556..738J}
  {556, 738}

\bibitem[\protect\citeauthoryear{{Jorstad} et~al.,}{{Jorstad}
  et~al.}{2005}]{jorstad_etal05}
{Jorstad} S.~G.,  et~al., 2005, \mn@doi [\aj] {10.1086/444593}, \href
  {http://adsabs.harvard.edu/abs/2005AJ....130.1418J} {130, 1418}

\bibitem[\protect\citeauthoryear{{Kadler}, {Ros}, {Lobanov}, {Falcke}  \&
  {Zensus}}{{Kadler} et~al.}{2004}]{kadler_etal04}
{Kadler} M.,  {Ros} E.,  {Lobanov} A.~P.,  {Falcke} H.,   {Zensus} J.~A.,
  2004, \mn@doi [\aap] {10.1051/0004-6361:20041051}, \href
  {http://adsabs.harvard.edu/abs/2004A%26A...426..481K} {426, 481}

\bibitem[\protect\citeauthoryear{{Karamanavis} et~al.,}{{Karamanavis}
  et~al.}{2016}]{karamanavis_etal16}
{Karamanavis} V.,  et~al., 2016, \mn@doi [\aap] {10.1051/0004-6361/201527225},
  \href {http://adsabs.harvard.edu/abs/2016A%26A...586A..60K} {586, A60}

\bibitem[\protect\citeauthoryear{{Kovalev} et~al.,}{{Kovalev}
  et~al.}{2005}]{kovalev_etal05}
{Kovalev} Y.~Y.,  et~al., 2005, \mn@doi [\aj] {10.1086/497430}, \href
  {http://adsabs.harvard.edu/cgi-bin/nph-bib_query?bibcode=2005AJ....130.2473K&db_key=AST}
  {130, 2473}

\bibitem[\protect\citeauthoryear{{Kovalev} et~al.,}{{Kovalev}
  et~al.}{2009}]{kovalev_etal09}
{Kovalev} Y.~Y.,  et~al., 2009, \mn@doi [\apjl] {10.1088/0004-637X/696/1/L17},
  \href {http://adsabs.harvard.edu/abs/2009ApJ...696L..17K} {696, L17}

\bibitem[\protect\citeauthoryear{{L{\"a}hteenm{\"a}ki} \&
  {Valtaoja}}{{L{\"a}hteenm{\"a}ki} \& {Valtaoja}}{1999}]{lv_99_iii}
{L{\"a}hteenm{\"a}ki} A.,  {Valtaoja} E.,  1999, \mn@doi [\apj]
  {10.1086/307587}, \href {http://esoads.eso.org/abs/1999ApJ...521..493L} {521,
  493}

\bibitem[\protect\citeauthoryear{{L{\"a}hteenm{\"a}ki} \&
  {Valtaoja}}{{L{\"a}hteenm{\"a}ki} \& {Valtaoja}}{2003}]{lv_03}
{L{\"a}hteenm{\"a}ki} A.,  {Valtaoja} E.,  2003, \mn@doi [\apj]
  {10.1086/374883}, \href {http://adsabs.harvard.edu/abs/2003ApJ...590...95L}
  {590, 95}

\bibitem[\protect\citeauthoryear{{L{\"a}hteenm{\"a}ki}, {Valtaoja}  \&
  {Wiik}}{{L{\"a}hteenm{\"a}ki} et~al.}{1999}]{lvw_99_ii}
{L{\"a}hteenm{\"a}ki} A.,  {Valtaoja} E.,   {Wiik} K.,  1999, \mn@doi [\apj]
  {10.1086/306649}, \href {http://adsabs.harvard.edu/abs/1999ApJ...511..112L}
  {511, 112}

\bibitem[\protect\citeauthoryear{{Lee}, {Lobanov}, {Krichbaum}, {Witzel},
  {Zensus}, {Bremer}, {Greve}  \& {Grewing}}{{Lee} et~al.}{2008}]{lee_etal08}
{Lee} S.-S.,  {Lobanov} A.~P.,  {Krichbaum} T.~P.,  {Witzel} A.,  {Zensus} A.,
  {Bremer} M.,  {Greve} A.,   {Grewing} M.,  2008, \mn@doi [\aj]
  {10.1088/0004-6256/136/1/159}, \href
  {http://adsabs.harvard.edu/abs/2008AJ....136..159L} {136, 159}

\bibitem[\protect\citeauthoryear{{Le{\'o}n-Tavares}, {Valtaoja}, {Tornikoski},
  {L{\"a}hteenm{\"a}ki}  \& {Nieppola}}{{Le{\'o}n-Tavares}
  et~al.}{2011}]{leon-tavares_etal11}
{Le{\'o}n-Tavares} J.,  {Valtaoja} E.,  {Tornikoski} M.,  {L{\"a}hteenm{\"a}ki}
  A.,   {Nieppola} E.,  2011, \mn@doi [\aap] {10.1051/0004-6361/201116664},
  \href {http://adsabs.harvard.edu/abs/2011A%26A...532A.146L} {532, A146}

\bibitem[\protect\citeauthoryear{{Leppanen}, {Zensus}  \& {Diamond}}{{Leppanen}
  et~al.}{1995}]{1995AJ....110.2479L}
{Leppanen} K.~J.,  {Zensus} J.~A.,   {Diamond} P.~J.,  1995, \mn@doi [\aj]
  {10.1086/117706}, \href {http://adsabs.harvard.edu/abs/1995AJ....110.2479L}
  {110, 2479}

\bibitem[\protect\citeauthoryear{{Lico} et~al.,}{{Lico}
  et~al.}{2014}]{lico_etal14}
{Lico} R.,  et~al., 2014, \mn@doi [\aap] {10.1051/0004-6361/201424341}, \href
  {http://adsabs.harvard.edu/abs/2014A%26A...571A..54L} {571, A54}

\bibitem[\protect\citeauthoryear{{Lister} \& {Homan}}{{Lister} \&
  {Homan}}{2005}]{lister_homan_05}
{Lister} M.~L.,  {Homan} D.~C.,  2005, \mn@doi [\aj] {10.1086/432969}, \href
  {http://adsabs.harvard.edu/abs/2005AJ....130.1389L} {130, 1389}

\bibitem[\protect\citeauthoryear{{Lister} et~al.,}{{Lister}
  et~al.}{2009}]{lister_etal09}
{Lister} M.~L.,  et~al., 2009, \mn@doi [\aj] {10.1088/0004-6256/137/3/3718},
  \href {http://adsabs.harvard.edu/abs/2009AJ....137.3718L} {137, 3718}

\bibitem[\protect\citeauthoryear{{Lister} et~al.,}{{Lister}
  et~al.}{2011}]{lister_etal11}
{Lister} M.~L.,  et~al., 2011, \mn@doi [\apj] {10.1088/0004-637X/742/1/27},
  \href {http://adsabs.harvard.edu/abs/2011ApJ...742...27L} {742, 27}

\bibitem[\protect\citeauthoryear{{Lister} et~al.,}{{Lister}
  et~al.}{2013}]{lister_etal13}
{Lister} M.~L.,  et~al., 2013, \mn@doi [\aj] {10.1088/0004-6256/146/5/120},
  \href {http://adsabs.harvard.edu/abs/2013AJ....146..120L} {146, 120}

\bibitem[\protect\citeauthoryear{{Lobanov}}{{Lobanov}}{1998}]{L98}
{Lobanov} A.~P.,  1998, \aap, \href
  {http://adsabs.harvard.edu/abs/1998A%26A...330...79L} {330, 79}

\bibitem[\protect\citeauthoryear{{Lobanov}}{{Lobanov}}{2005}]{lobanov_05}
{Lobanov} A.~P.,  2005, preprint, \href
  {http://adsabs.harvard.edu/abs/2005astro.ph..3225L} {} (\mn@eprint {arXiv}
  {astro-ph/0503225})

\bibitem[\protect\citeauthoryear{{Marcaide} \& {Shapiro}}{{Marcaide} \&
  {Shapiro}}{1984}]{marcaide_shapiro_84}
{Marcaide} J.~M.,  {Shapiro} I.~I.,  1984, \mn@doi [\apj] {10.1086/161592},
  \href {http://adsabs.harvard.edu/abs/1984ApJ...276...56M} {276, 56}

\bibitem[\protect\citeauthoryear{{Marscher}}{{Marscher}}{2008}]{marscher08}
{Marscher} A.~P.,  2008, in {Rector} T.~A.,  {De Young} D.~S.,  eds,
  Astronomical Society of the Pacific Conference Series Vol. 386, Extragalactic
  Jets: Theory and Observation from Radio to Gamma Ray. p.~437

\bibitem[\protect\citeauthoryear{{Marscher} \& {Gear}}{{Marscher} \&
  {Gear}}{1985}]{marscher_gear_85}
{Marscher} A.~P.,  {Gear} W.~K.,  1985, \mn@doi [\apj] {10.1086/163592}, \href
  {http://adsabs.harvard.edu/abs/1985ApJ...298..114M} {298, 114}

\bibitem[\protect\citeauthoryear{{Marscher} et~al.,}{{Marscher}
  et~al.}{2008}]{marscher_etal08}
{Marscher} A.~P.,  et~al., 2008, \mn@doi [\nat] {10.1038/nature06895}, \href
  {http://adsabs.harvard.edu/abs/2008Natur.452..966M} {452, 966}

\bibitem[\protect\citeauthoryear{{Marscher} et~al.,}{{Marscher}
  et~al.}{2010a}]{marscher_etal10a}
{Marscher} A.~P.,  et~al., 2010a, preprint, \href
  {http://adsabs.harvard.edu/abs/2010arXiv1002.0806M} {} (\mn@eprint {arXiv}
  {1002.0806})

\bibitem[\protect\citeauthoryear{{Marscher} et~al.,}{{Marscher}
  et~al.}{2010b}]{marscher_etal10b}
{Marscher} A.~P.,  et~al., 2010b, \mn@doi [\apjl]
  {10.1088/2041-8205/710/2/L126}, \href
  {http://adsabs.harvard.edu/abs/2010ApJ...710L.126M} {710, L126}

\bibitem[\protect\citeauthoryear{{Mattox} et~al.,}{{Mattox}
  et~al.}{1996}]{Mattox_etal96}
{Mattox} J.~R.,  et~al., 1996, \mn@doi [\apj] {10.1086/177068}, \href
  {http://esoads.eso.org/abs/1996ApJ...461..396M} {461, 396}

\bibitem[\protect\citeauthoryear{{Morozova} et~al.,}{{Morozova}
  et~al.}{2014}]{morozova_etal14}
{Morozova} D.~A.,  et~al., 2014, \mn@doi [\aj] {10.1088/0004-6256/148/3/42},
  \href {http://adsabs.harvard.edu/abs/2014AJ....148...42M} {148, 42}

\bibitem[\protect\citeauthoryear{{O'Sullivan} \& {Gabuzda}}{{O'Sullivan} \&
  {Gabuzda}}{2009a}]{og_09mf}
{O'Sullivan} S.~P.,  {Gabuzda} D.~C.,  2009a, \mn@doi [\mnras]
  {10.1111/j.1365-2966.2008.14213.x}, \href
  {http://esoads.eso.org/abs/2009MNRAS.393..429O} {393, 429}

\bibitem[\protect\citeauthoryear{{O'Sullivan} \& {Gabuzda}}{{O'Sullivan} \&
  {Gabuzda}}{2009b}]{og_09cs}
{O'Sullivan} S.~P.,  {Gabuzda} D.~C.,  2009b, \mn@doi [\mnras]
  {10.1111/j.1365-2966.2009.15428.x}, \href
  {http://adsabs.harvard.edu/abs/2009MNRAS.400...26O} {400, 26}

\bibitem[\protect\citeauthoryear{{Orienti}, {Venturi}, {Dallacasa},
  {D'Ammando}, {Giroletti}, {Giovannini}, {Vercellone}  \& {Tavani}}{{Orienti}
  et~al.}{2011}]{2011MNRAS.417..359O}
{Orienti} M.,  {Venturi} T.,  {Dallacasa} D.,  {D'Ammando} F.,  {Giroletti} M.,
   {Giovannini} G.,  {Vercellone} S.,   {Tavani} M.,  2011, \mn@doi [\mnras]
  {10.1111/j.1365-2966.2011.19272.x}, \href
  {http://adsabs.harvard.edu/abs/2011MNRAS.417..359O} {417, 359}

\bibitem[\protect\citeauthoryear{{Pacholczyk} \& {Swihart}}{{Pacholczyk} \&
  {Swihart}}{1967}]{PS67}
{Pacholczyk} A.~G.,  {Swihart} T.~L.,  1967, \mn@doi [\apj] {10.1086/149364},
  \href {http://esoads.eso.org/abs/1967ApJ...150..647P} {150, 647}

\bibitem[\protect\citeauthoryear{{Piner} \& {Kingham}}{{Piner} \&
  {Kingham}}{1998}]{pk_98}
{Piner} B.~G.,  {Kingham} K.~A.,  1998, \mn@doi [\apj] {10.1086/306346}, \href
  {http://adsabs.harvard.edu/abs/1998ApJ...507..706P} {507, 706}

\bibitem[\protect\citeauthoryear{{Planck Collaboration} et~al.,}{{Planck
  Collaboration} et~al.}{2015}]{planck_15}
{Planck Collaboration} et~al., 2015, preprint, \href
  {http://adsabs.harvard.edu/abs/2015arXiv150201589P} {} (\mn@eprint {arXiv}
  {1502.01589})

\bibitem[\protect\citeauthoryear{{Pushkarev} \& {Kovalev}}{{Pushkarev} \&
  {Kovalev}}{2012}]{pushkarev_kovalev_12}
{Pushkarev} A.~B.,  {Kovalev} Y.~Y.,  2012, \mn@doi [\aap]
  {10.1051/0004-6361/201219352}, \href
  {http://adsabs.harvard.edu/abs/2012A%26A...544A..34P} {544, A34}

\bibitem[\protect\citeauthoryear{{Pushkarev}, {Kovalev}, {Lister}  \&
  {Savolainen}}{{Pushkarev} et~al.}{2009}]{2009AA...507L..33P}
{Pushkarev} A.~B.,  {Kovalev} Y.~Y.,  {Lister} M.~L.,   {Savolainen} T.,  2009,
  \mn@doi [\aap] {10.1051/0004-6361/200913422}, \href
  {http://adsabs.harvard.edu/abs/2009A%26A...507L..33P} {507, L33}

\bibitem[\protect\citeauthoryear{{Pushkarev}, {Kovalev}  \&
  {Lister}}{{Pushkarev} et~al.}{2010}]{pushkarev_etal10}
{Pushkarev} A.~B.,  {Kovalev} Y.~Y.,   {Lister} M.~L.,  2010, \mn@doi [\apjl]
  {10.1088/2041-8205/722/1/L7}, \href
  {http://adsabs.harvard.edu/abs/2010ApJ...722L...7P} {722, L7}

\bibitem[\protect\citeauthoryear{{Pushkarev}, {Hovatta}, {Kovalev}, {Lister},
  {Lobanov}, {Savolainen}  \& {Zensus}}{{Pushkarev}
  et~al.}{2012}]{pushkarev_etal12}
{Pushkarev} A.~B.,  {Hovatta} T.,  {Kovalev} Y.~Y.,  {Lister} M.~L.,  {Lobanov}
  A.~P.,  {Savolainen} T.,   {Zensus} J.~A.,  2012, \mn@doi [\aap]
  {10.1051/0004-6361/201219173}, \href
  {http://adsabs.harvard.edu/abs/2012A%26A...545A.113P} {545, A113}

\bibitem[\protect\citeauthoryear{{Ramakrishnan} et~al.,}{{Ramakrishnan}
  et~al.}{2014}]{ramakrishnan_etal14}
{Ramakrishnan} V.,  et~al., 2014, \mn@doi [\mnras] {10.1093/mnras/stu1873},
  \href {http://adsabs.harvard.edu/abs/2014MNRAS.445.1636R} {445, 1636}

\bibitem[\protect\citeauthoryear{{Ramakrishnan}, {Hovatta}, {Nieppola},
  {Tornikoski}, {L{\"a}hteenm{\"a}ki}  \& {Valtaoja}}{{Ramakrishnan}
  et~al.}{2015}]{ramakrishnan_etal15}
{Ramakrishnan} V.,  {Hovatta} T.,  {Nieppola} E.,  {Tornikoski} M.,
  {L{\"a}hteenm{\"a}ki} A.,   {Valtaoja} E.,  2015, \mn@doi [\mnras]
  {10.1093/mnras/stv321}, \href
  {http://adsabs.harvard.edu/abs/2015MNRAS.452.1280R} {452, 1280}

\bibitem[\protect\citeauthoryear{{Rani}, {Krichbaum}, {Marscher}, {Jorstad},
  {Hodgson}, {Fuhrmann}  \& {Zensus}}{{Rani} et~al.}{2014}]{rani_etal14}
{Rani} B.,  {Krichbaum} T.~P.,  {Marscher} A.~P.,  {Jorstad} S.~G.,  {Hodgson}
  J.~A.,  {Fuhrmann} L.,   {Zensus} J.~A.,  2014, \mn@doi [\aap]
  {10.1051/0004-6361/201424796}, \href
  {http://adsabs.harvard.edu/abs/2014A%26A...571L...2R} {571, L2}

\bibitem[\protect\citeauthoryear{{Readhead}}{{Readhead}}{1994}]{readhead_94}
{Readhead} A.~C.~S.,  1994, \mn@doi [\apj] {10.1086/174038}, \href
  {http://adsabs.harvard.edu/abs/1994ApJ...426...51R} {426, 51}

\bibitem[\protect\citeauthoryear{{Richards} et~al.,}{{Richards}
  et~al.}{2011}]{richards_etal11}
{Richards} J.~L.,  et~al., 2011, \mn@doi [\apjs] {10.1088/0067-0049/194/2/29},
  \href {http://esoads.eso.org/abs/2011ApJS..194...29R} {194, 29}

\bibitem[\protect\citeauthoryear{{Roberts}, {Wardle}  \& {Brown}}{{Roberts}
  et~al.}{1994}]{roberts_etal94}
{Roberts} D.~H.,  {Wardle} J.~F.~C.,   {Brown} L.~F.,  1994, \mn@doi [\apj]
  {10.1086/174180}, \href {http://adsabs.harvard.edu/abs/1994ApJ...427..718R}
  {427, 718}

\bibitem[\protect\citeauthoryear{{Savolainen}, {Wiik}, {Valtaoja}, {Jorstad}
  \& {Marscher}}{{Savolainen} et~al.}{2002}]{savolainen_etal02}
{Savolainen} T.,  {Wiik} K.,  {Valtaoja} E.,  {Jorstad} S.~G.,   {Marscher}
  A.~P.,  2002, \mn@doi [\aap] {10.1051/0004-6361:20021236}, \href
  {http://esoads.eso.org/abs/2002A%26A...394..851S} {394, 851}

\bibitem[\protect\citeauthoryear{{Savolainen}, {Homan}, {Hovatta}, {Kadler},
  {Kovalev}, {Lister}, {Ros}  \& {Zensus}}{{Savolainen}
  et~al.}{2010}]{savolainen_etal10}
{Savolainen} T.,  {Homan} D.~C.,  {Hovatta} T.,  {Kadler} M.,  {Kovalev} Y.~Y.,
   {Lister} M.~L.,  {Ros} E.,   {Zensus} J.~A.,  2010, \mn@doi [\aap]
  {10.1051/0004-6361/200913740}, \href
  {http://adsabs.harvard.edu/abs/2010A%26A...512A..24S} {512, A24}

\bibitem[\protect\citeauthoryear{{Schneider} et~al.,}{{Schneider}
  et~al.}{2010}]{schneider_etal10}
{Schneider} D.~P.,  et~al., 2010, \mn@doi [\aj] {10.1088/0004-6256/139/6/2360},
  \href {http://adsabs.harvard.edu/abs/2010AJ....139.2360S} {139, 2360}

\bibitem[\protect\citeauthoryear{{Shepherd}}{{Shepherd}}{1997}]{shepherd_97}
{Shepherd} M.~C.,  1997, in {Hunt} G.,  {Payne} H.,  eds,  Astronomical Society
  of the Pacific Conference Series Vol. 125, Astronomical Data Analysis
  Software and Systems VI. p.~77

\bibitem[\protect\citeauthoryear{{Shepherd}, {Pearson}  \& {Taylor}}{{Shepherd}
  et~al.}{1994}]{shepherd_etal94}
{Shepherd} M.~C.,  {Pearson} T.~J.,   {Taylor} G.~B.,  1994, in Bulletin of the
  American Astronomical Society. pp 987--989

\bibitem[\protect\citeauthoryear{{Smith} \& {Bechetti}}{{Smith} \&
  {Bechetti}}{2010}]{smith_atel10}
{Smith} P.~S.,  {Bechetti} G.~J.,  2010, The Astronomer's Telegram, \href
  {http://adsabs.harvard.edu/abs/2010ATel.2623....1S} {2623}

\bibitem[\protect\citeauthoryear{{Sokoloff}, {Bykov}, {Shukurov},
  {Berkhuijsen}, {Beck}  \& {Poezd}}{{Sokoloff} et~al.}{1998}]{sokoloff_etal98}
{Sokoloff} D.~D.,  {Bykov} A.~A.,  {Shukurov} A.,  {Berkhuijsen} E.~M.,  {Beck}
  R.,   {Poezd} A.~D.,  1998, \mn@doi [\mnras]
  {10.1046/j.1365-8711.1998.01782.x}, \href
  {http://adsabs.harvard.edu/abs/1998MNRAS.299..189S} {299, 189}

\bibitem[\protect\citeauthoryear{{Sokolovsky}, {Kovalev}, {Pushkarev}  \&
  {Lobanov}}{{Sokolovsky} et~al.}{2011}]{sokolovsky_etal11}
{Sokolovsky} K.~V.,  {Kovalev} Y.~Y.,  {Pushkarev} A.~B.,   {Lobanov} A.~P.,
  2011, \mn@doi [\aap] {10.1051/0004-6361/201016072}, \href
  {http://esoads.eso.org/abs/2011A%26A...532A..38S} {532, A38}

\bibitem[\protect\citeauthoryear{{Stirling} et~al.,}{{Stirling}
  et~al.}{2003}]{stirling_etal03}
{Stirling} A.~M.,  et~al., 2003, \mn@doi [\mnras]
  {10.1046/j.1365-8711.2003.06448.x}, \href
  {http://adsabs.harvard.edu/abs/2003MNRAS.341..405S} {341, 405}

\bibitem[\protect\citeauthoryear{{Tavecchio}, {Ghisellini}, {Bonnoli}  \&
  {Ghirlanda}}{{Tavecchio} et~al.}{2010}]{tavecchio_etal10}
{Tavecchio} F.,  {Ghisellini} G.,  {Bonnoli} G.,   {Ghirlanda} G.,  2010,
  \mn@doi [\mnras] {10.1111/j.1745-3933.2010.00867.x}, \href
  {http://adsabs.harvard.edu/abs/2010MNRAS.405L..94T} {405, L94}

\bibitem[\protect\citeauthoryear{{Taylor}}{{Taylor}}{2000}]{taylor_00}
{Taylor} G.~B.,  2000, \mn@doi [\apj] {10.1086/308666}, \href
  {http://adsabs.harvard.edu/abs/2000ApJ...533...95T} {533, 95}

\bibitem[\protect\citeauthoryear{{Taylor} \& {Myers}}{{Taylor} \&
  {Myers}}{2000}]{taylor_myers_00}
{Taylor} G.~B.,  {Myers} S.~T.,  2000, VLBA Scientific Memo, 26

\bibitem[\protect\citeauthoryear{{Taylor}, {Stil}  \& {Sunstrum}}{{Taylor}
  et~al.}{2009}]{taylor_etal09}
{Taylor} A.~R.,  {Stil} J.~M.,   {Sunstrum} C.,  2009, \mn@doi [\apj]
  {10.1088/0004-637X/702/2/1230}, \href
  {http://adsabs.harvard.edu/abs/2009ApJ...702.1230T} {702, 1230}

\bibitem[\protect\citeauthoryear{{Tribble}}{{Tribble}}{1991}]{T91}
{Tribble} P.~C.,  1991, \mnras, \href
  {http://esoads.eso.org/abs/1991MNRAS.250..726T} {250, 726}

\bibitem[\protect\citeauthoryear{{Valtaoja} \& {Teraesranta}}{{Valtaoja} \&
  {Teraesranta}}{1996}]{vt_96}
{Valtaoja} E.,  {Teraesranta} H.,  1996, \aaps, \href
  {http://adsabs.harvard.edu/abs/1996A%26AS..120C.491V} {120, C491}

\bibitem[\protect\citeauthoryear{{Valtaoja}, {L{\"a}hteenm{\"a}ki},
  {Ter{\"a}sranta}  \& {Lainela}}{{Valtaoja} et~al.}{1999}]{valtaoja_etal99}
{Valtaoja} E.,  {L{\"a}hteenm{\"a}ki} A.,  {Ter{\"a}sranta} H.,   {Lainela} M.,
   1999, \mn@doi [\apjs] {10.1086/313170}, \href
  {http://adsabs.harvard.edu/abs/1999ApJS..120...95V} {120, 95}

\bibitem[\protect\citeauthoryear{{Walker}, {Dhawan}, {Romney}, {Kellermann}  \&
  {Vermeulen}}{{Walker} et~al.}{2000}]{walker_etal2000}
{Walker} R.~C.,  {Dhawan} V.,  {Romney} J.~D.,  {Kellermann} K.~I.,
  {Vermeulen} R.~C.,  2000, \mn@doi [\apj] {10.1086/308372}, \href
  {http://esoads.eso.org/abs/2000ApJ...530..233W} {530, 233}

\bibitem[\protect\citeauthoryear{{Welsh}}{{Welsh}}{1999}]{welsh_99}
{Welsh} W.~F.,  1999, \mn@doi [\pasp] {10.1086/316457}, \href
  {http://adsabs.harvard.edu/abs/1999PASP..111.1347W} {111, 1347}

\bibitem[\protect\citeauthoryear{{Wright}}{{Wright}}{2006}]{wright_06}
{Wright} E.~L.,  2006, \mn@doi [\pasp] {10.1086/510102}, \href
  {http://esoads.eso.org/abs/2006PASP..118.1711W} {118, 1711}

\bibitem[\protect\citeauthoryear{{Zamaninasab}, {Clausen-Brown}, {Savolainen}
  \& {Tchekhovskoy}}{{Zamaninasab} et~al.}{2014}]{zamaninasab_etal14}
{Zamaninasab} M.,  {Clausen-Brown} E.,  {Savolainen} T.,   {Tchekhovskoy} A.,
  2014, \mn@doi [\nat] {10.1038/nature13399}, \href
  {http://adsabs.harvard.edu/abs/2014Natur.510..126Z} {510, 126}

\bibitem[\protect\citeauthoryear{{Zavala} \& {Taylor}}{{Zavala} \&
  {Taylor}}{2001}]{zt_01}
{Zavala} R.~T.,  {Taylor} G.~B.,  2001, \mn@doi [\apjl] {10.1086/319653}, \href
  {http://adsabs.harvard.edu/abs/2001ApJ...550L.147Z} {550, L147}

\bibitem[\protect\citeauthoryear{{Zdziarski}, {Sikora}, {Pjanka}  \&
  {Tchekhovskoy}}{{Zdziarski} et~al.}{2015}]{zdziarski_etal15}
{Zdziarski} A.~A.,  {Sikora} M.,  {Pjanka} P.,   {Tchekhovskoy} A.,  2015,
  \mn@doi [\mnras] {10.1093/mnras/stv986}, \href
  {http://adsabs.harvard.edu/abs/2015MNRAS.451..927Z} {451, 927}

\makeatother
\end{thebibliography}


\appendix
\section{Supporting information}
Supplementary material to go in online version of the article:
\begin{enumerate}
\item Table \ref{tbo_frq}. VLBA central frequencies.
\item Table \ref{tbo_ampc}. Amplitude corrections for the S2087E experiment.
\item Table \ref{tbo_mdf}. Results of Gaussian model fitting and component parameters at 4.6--43.2~GHz
\end{enumerate}

\begin{table}
\caption{VLBA central frequencies.\label{tbo_frq}}
\begin{tabular}{lcr}
\hline
IEEE band & IF & Frequency\\
 & &(MHz)\\
\hline
C & 1 & 4604.5\\
  & 2 & 4612.5\\
  & 3 & 4999.9\\
  & 4 & 5007.5\\
X & 1 & 8104.5\\
  & 2 & 8112.5\\
  & 3 & 8425.5\\
  & 4 & 8433.5\\
K$_u$ & 1 & 15353.5\\
  & 2 & 15361.5\\
  & 3 & 15369.5\\
  & 4 & 15377.5\\
K & 1 & 23792.5\\
  & 2 & 23780.5\\
  & 3 & 23808.5\\
  & 4 & 23816.5\\
Q & 1 & 43205.5\\
 & 2 & 43211.5\\
 & 3 & 43221.5\\
 & 4 & 43229.5\\
\hline
\end{tabular}
\end{table}


\begin{table}
 \caption{Amplitude corrections for the S2087E VLBA experiment.\label{tbo_ampc}}
  \begin{tabular}{@{}cccccc@{}}
\hline
Antenna & Band & Epoch & IF & Polarization & Correction\\
(1) & (2) & (3) & (4) & (5) & (6)\\
\hline
BR & C & All & 1,2 & RCP & 1.10\\
KP & C & All & 1,2 & RCP & 1.10\\
KP & C & All & 1,2 & LCP & 1.13\\ 
MK & C & All & 1 & LCP & 1.10\\
PT & C & All & 4 & RCP & 0.90\\
FD & X & All & 1,2 & RCP & 1.08\\
KP & X & All & 1,2 & RCP & 0.89\\
OV & X & 3,4 & 1 & LCP & 1.12\\
SC & K$_u$ & 1 & 1$-$4 & RCP, LCP & 0.86\\
BR & K & All & 1$-$4 & LCP, RCP & 0.76\\
HN & K & 2,3 & 1$-$4 & LCP, RCP & 0.84\\
KP & K & All & 1$-$4 & RCP & 1.33\\
PT & K & All & 3,4 & LCP, RCP & 1.11\\
HN & Q & 2 & 1$-$4 & LCP, RCP & 1.18\\
NL & Q & 1 & 1$-$4 & LCP, RCP & 1.18\\
OV & Q & 3 & 1$-$4 & RCP & 1.13\\
PT & Q & 2 & 2,4 & LCP & 0.89 \\
SC & Q & All & 1$-$4 & LCP, RCP & 0.88\\
\hline
\end{tabular}

\medskip
Column designation: (1) antenna name; (2) radio band name; (3) observation epoch (epochs are labeled as follows: 1 for 2010--05--24, 2 for 2010--07--09, 3 for 2010--08--28, 4 for 2010--10--18); (4) number of the frequency channel (IF); (5) polarization (right of left circular); (6) amplitude correction coefficient.
\end{table}

\onecolumn
\begin{table}
 \begin{minipage}{168mm}
  \caption{Results of Gaussian model fitting and component parameters at 4.6--43.2~GHz.\label{tbo_mdf}}
  \begin{tabular}{@{}lcr@{$\pm$}lr@{$\pm$}lcccc@{}}
\hline
Date & Name & \multicolumn{2}{c}{Flux density} & \multicolumn{2}{c}{Distance} & P.A. & Major & Ratio & Major P.A.\\
 &  & \multicolumn{2}{c}{(Jy)} & \multicolumn{2}{c}{(mas)} &($^\circ$)& (mas) & & ($^\circ$) \\
(1)& (2) & \multicolumn{2}{c}{(3)} & \multicolumn{2}{c}{(4)} &(5)& (6) &(7) & (8) \\
\hline
\multicolumn{10}{c}{4.6~GHz}\\
\hline
2010--05--24
& Core & 0.147&0.007&0.00&0.04&0.0&0.202&1.0&\dots\\
& C2 & 0.055&0.003&1.02&0.07&173.8&0.480&1.0&\dots\\
& C1 & 0.0263&0.0013&3.02&0.14&166.0&1.094&1.0&\dots\\
& U1 & 0.0210&0.0011&4.43&0.18&162.9&1.918&1.0&\dots\\
& U0 & 0.0126&0.0006&7.5&0.4&$-$177.6&4.607&1.0&\dots\\
2010--07--09
& Core & 0.147&0.007&0.00&0.04&0.0&0.146&1.0&\dots\\
& C2 & 0.055&0.003&1.03&0.06&171.8&0.508&1.0&\dots\\
& C1 & 0.0265&0.0013&3.13&0.11&166.1&1.102&1.0&\dots\\
& U1 & 0.0184&0.0009&4.62&0.19&161.1&1.906&1.0&\dots\\
& U0 & 0.0116&0.0006&7.4&0.5&$-$178.6&4.436&1.0&\dots\\
2010--08--28
& Core & 0.168&0.008&0.00&0.04&0.0&0.097&1.0&\dots\\
& C2 & 0.058&0.003&1.03&0.07&170.4&0.546&1.0&\dots\\
& C1 & 0.0241&0.0012&3.04&0.14&165.4&1.085&1.0&\dots\\
& U1 & 0.0222&0.0011&4.6&0.2&163.1&2.138&1.0&\dots\\
& U0 & 0.0129&0.0007&7.8&0.6&$-$176.4&6.348&1.0&\dots\\
2010--10--18
& Core & 0.179&0.009&0.00&0.03&0.0&0.105&1.0&\dots\\
& C2 & 0.063&0.003&1.03&0.07&170.5&0.530&1.0&\dots\\
& C1 & 0.0279&0.0014&3.15&0.14&166.0&1.141&1.0&\dots\\
& U1 & 0.0192&0.0010&4.8&0.3&161.5&2.148&1.0&\dots\\
& U0 & 0.0070&0.0005&8.2&0.6&$-$173.6&3.341&1.0&\dots\\
\hline
\multicolumn{10}{c}{5.0~GHz}\\
\hline
2010--05--24
& Core & 0.148&0.007&0.00&0.03&0.0&0.083&1.0&\dots\\
& C2 & 0.053&0.003&1.01&0.06&171.5&0.497&1.0&\dots\\
& C1 & 0.0288&0.0014&3.19&0.11&165.6&1.227&1.0&\dots\\
& U1 & 0.0152&0.0008&4.8&0.2&162.3&2.017&1.0&\dots\\
& U0 & 0.0113&0.0006&8.0&0.4&$-$177.0&5.201&1.0&\dots\\
2010--07--09
& Core & 0.153&0.008&0.00&0.04&0.0&0.160&1.0&\dots\\
& C2 & 0.050&0.003&1.03&0.07&170.3&0.444&1.0&\dots\\
& C1 & 0.0299&0.0015&3.13&0.14&166.0&1.324&1.0&\dots\\
& U1 & 0.0145&0.0008&4.9&0.2&161.2&1.899&1.0&\dots\\
& U0 & 0.0102&0.0005&7.8&0.4&$-$173.8&4.755&1.0&\dots\\
2010--08--28
& Core & 0.172&0.009&0.00&0.03&0.0&0.155&1.0&\dots\\
& C2 & 0.055&0.003&1.02&0.06&172.4&0.534&1.0&\dots\\
& C1 & 0.0263&0.0013&3.07&0.12&165.4&1.235&1.0&\dots\\
& U1 & 0.0180&0.0009&4.71&0.18&163.2&1.903&1.0&\dots\\
& U0 & 0.0098&0.0005&7.8&0.4&$-$176.8&4.938&1.0&\dots\\
2010--10--18
& Core & 0.180&0.009&0.00&0.03&0.0&0.151&1.0&\dots\\
& C2 & 0.059&0.003&1.00&0.06&170.7&0.502&1.0&\dots\\
& C1 & 0.0241&0.0012&3.04&0.14&165.5&1.206&1.0&\dots\\
& U1 & 0.0207&0.0011&4.6&0.2&163.3&2.264&1.0&\dots\\
& U0 & 0.0085&0.0005&8.2&0.5&$-$174.5&4.010&1.0&\dots\\
 \hline
\end{tabular}

\medskip
Column designation: (1) observation date; (2) the name of the component ("U" stands for unclassified components); (3) the integrated flux density in the component and its error; (4) the radial distance of the component center from the center of the map and its error; (5) the position angle of the center of the component relative to the map center; (6) the FWHM major axis of the fitted Gaussian; (7) axial ratio of major and minor axes of fitted Gaussian; (8) major axis position angle of fitted Gaussian.
\end{minipage}
\end{table}

\begin{table}
 \centering
 \begin{minipage}{168mm}
  \contcaption{}
  \begin{tabular}{@{}lcr@{$\pm$}lr@{$\pm$}lcccc@{}}
\hline
Date & Name & \multicolumn{2}{c}{Flux density} & \multicolumn{2}{c}{Distance} & P.A. & Major & Ratio & Major P.A.\\
 &  & \multicolumn{2}{c}{(Jy)} & \multicolumn{2}{c}{(mas)} &($^\circ$)& (mas) & & ($^\circ$) \\
(1)& (2) & \multicolumn{2}{c}{(3)} & \multicolumn{2}{c}{(4)} &(5)& (6) &(7) & (8) \\
\hline
\multicolumn{10}{c}{8.1~GHz}\\
\hline
2010--05--24
& Core & 0.214&0.011&0.00&0.02&0.0&0.204&0.4&$-$17.8\\
& C2 & 0.0365&0.0018&1.04&0.07&172.0&0.515&1.0&\dots\\
& C1 & 0.0186&0.0010&3.07&0.14&166.0&1.176&1.0&\dots\\
& U1 & 0.0161&0.0008&4.60&0.16&162.6&2.056&1.0&\dots\\
& U0 & 0.0053&0.0003&8.3&0.4&$-$173.8&2.955&1.0&\dots\\
2010--07--09
& Core & 0.210&0.011&0.00&0.02&0.0&0.200&0.2&2.0\\
& C2 & 0.038&0.002&1.01&0.08&171.7&0.514&1.0&\dots\\
& C1 & 0.0211&0.0011&3.17&0.13&165.9&1.213&1.0&\dots\\
& U1 & 0.0112&0.0006&4.97&0.19&161.6&2.069&1.0&\dots\\
& U0 & 0.0059&0.0003&7.9&0.4&$-$172.9&3.712&1.0&\dots\\
2010--08--28
& Core & 0.231&0.012&0.00&0.02&0.0&0.181&0.5&0.3\\
& C2 & 0.040&0.002&1.03&0.07&172.2&0.469&1.0&\dots\\
& C1 & 0.0184&0.0010&3.12&0.14&166.5&1.195&1.0&\dots\\
& U1 & 0.0142&0.0007&4.7&0.2&162.3&2.229&1.0&\dots\\
& U0 & 0.0041&0.0003&8.1&0.4&$-$175.5&2.839&1.0&\dots\\
2010--10--18
& Core & 0.264&0.013&0.00&0.02&0.0&0.171&0.4&$-$1.6\\
& C2 & 0.040&0.002&1.03&0.07&170.8&0.461&1.0&\dots\\
& C1 & 0.0177&0.0009&3.10&0.13&166.2&1.173&1.0&\dots\\
& U1 & 0.0125&0.0007&4.71&0.19&161.9&1.917&1.0&\dots\\
& U0 & 0.0069&0.0004&7.7&0.3&$-$172.1&3.648&1.0&\dots\\
\hline
\multicolumn{10}{c}{8.4~GHz}\\
\hline
2010--05--24
& Core & 0.222&0.011&0.00&0.02&0.0&0.170&0.4&$-$4.8\\
& C2 & 0.0345&0.0017&1.04&0.07&172.1&0.466&1.0&\dots\\
& C1 & 0.0207&0.0011&3.18&0.13&165.9&1.223&1.0&\dots\\
& U1 & 0.0130&0.0007&4.81&0.18&164.1&2.079&1.0&\dots\\
& U0 & 0.0048&0.0003&8.7&0.5&$-$177.4&4.299&1.0&\dots\\
2010--07--09
& Core & 0.218&0.011&0.00&0.02&0.0&0.194&0.4&0.7\\
& C2 & 0.0367&0.0019&1.02&0.06&172.0&0.457&1.0&\dots\\
& C1 & 0.0211&0.0011&3.16&0.13&165.9&1.249&1.0&\dots\\
& U1 & 0.0125&0.0006&4.86&0.19&161.6&2.304&1.0&\dots\\
& U0 & 0.0050&0.0003&8.9&0.4&$-$172.8&3.754&1.0&\dots\\
2010--08--28
& Core & 0.237&0.012&0.00&0.02&0.0&0.168&0.4&$-$4.3\\
& C2 & 0.0380&0.0019&1.03&0.06&171.8&0.453&1.0&\dots\\
& C1 & 0.0201&0.0010&3.08&0.13&165.6&1.301&1.0&\dots\\
& U1 & 0.0102&0.0005&4.68&0.19&161.8&1.807&1.0&\dots\\
& U0 & 0.0099&0.0005&7.0&0.4&177.0&5.050&1.0&\dots\\
2010--10--18
& Core & 0.272&0.014&0.00&0.01&0.0&0.177&0.4&$-$4.2\\
& C2 & 0.039&0.002&1.02&0.06&171.0&0.448&1.0&\dots\\
& C1 & 0.0191&0.0010&3.13&0.13&166.2&1.231&1.0&\dots\\
& U1 & 0.0123&0.0006&4.84&0.16&162.0&1.953&1.0&\dots\\
& U0 & 0.0061&0.0003&8.1&0.5&$-$174.1&4.883&1.0&\dots\\
\hline
\end{tabular}
\end{minipage}
\end{table}

\begin{table}
 \centering
 \begin{minipage}{168mm}
  \contcaption{}
  \begin{tabular}{@{}lcr@{$\pm$}lr@{$\pm$}lcccc@{}}
\hline
Date & Name & \multicolumn{2}{c}{Flux density} & \multicolumn{2}{c}{Distance} & P.A. & Major & Ratio & Major P.A.\\
 &  & \multicolumn{2}{c}{(Jy)} & \multicolumn{2}{c}{(mas)} &($^\circ$)& (mas) & & ($^\circ$) \\
(1)& (2) & \multicolumn{2}{c}{(3)} & \multicolumn{2}{c}{(4)} &(5)& (6) &(7) & (8) \\
\hline
\multicolumn{10}{c}{15.4~GHz}\\
\hline
2009--06--25
& Core & 0.267&0.013&0.00&0.03&0.0&0.099&1.0&\dots\\
& C2 & 0.0235&0.0019&0.94&0.17&173.2&0.700&1.0&\dots\\
& C1 & 0.0216&0.0014&3.5&0.2&165.6&1.565&1.0&\dots\\
2009--12--26
& Core & 0.285&0.014&0.00&0.02&0.0&0.110&0.3&$-$7.6\\
& C2 & 0.0220&0.0015&0.97&0.15&171.9&0.670&1.0&\dots\\
& C1 & 0.0197&0.0013&3.6&0.3&166.2&1.821&1.0&\dots\\
2010--05--24
& Core & 0.282&0.014&0.00&0.015&0.0&0.119&0.4&$-$1.8\\
& C2 & 0.0229&0.0013&0.98&0.09&172.2&0.508&1.0&\dots\\
& C1 & 0.0209&0.0012&3.4&0.2&164.2&1.819&1.0&\dots\\
2010--07--09
& Core & 0.286&0.014&0.000&0.014&0.0&0.120&0.4&7.2\\
& C2 & 0.0238&0.0014&0.98&0.09&172.5&0.520&1.0&\dots\\
& C1 & 0.0209&0.0012&3.6&0.2&166.3&1.944&1.0&\dots\\
2010--08--28
& Core & 0.324&0.016&0.000&0.011&0.0&0.131&0.2&7.0\\
& C2 & 0.0232&0.0013&1.01&0.07&172.2&0.488&1.0&\dots\\
& C1 & 0.0196&0.0011&3.31&0.19&163.8&1.860&1.0&\dots\\
2010--10--18
& Core & 0.41&0.02&0.000&0.012&0.0&0.130&0.3&8.2\\
& C2 & 0.0249&0.0014&1.013&0.08&171.5&0.494&1.0&\dots\\
& C1 & 0.0186&0.0011&3.5&0.2&165.8&1.854&1.0&\dots\\
2010--12--24
& Core & 0.371&0.019&0.000&0.011&0.0&0.026&1.0&\dots\\
& C3 & 0.151&0.008&0.124&0.017&$-$173.2&0.041&1.0&\dots\\
& C2 & 0.0215&0.0012&1.10&0.08&170.9&0.455&1.0&\dots\\
& C1 & 0.0176&0.0011&3.6&0.2&166.8&1.989&1.0&\dots\\
2011--04--11
& Core & 0.45&0.02&0.000&0.010&0.0&0.023&1.0&\dots\\
& C3 & 0.175&0.009&0.147&0.017&$-$175.1&0.084&1.0&\dots\\
& C2 & 0.0260&0.0014&1.12&0.07&170.2&0.431&1.0&\dots\\
& C1 & 0.0173&0.0011&3.4&0.2&168.8&1.759&1.0&\dots\\
2011--05--26
& Core & 0.44&0.02&0.000&0.014&0.0&0.031&1.0&\dots\\
& C3 & 0.145&0.007&0.17&0.03&$-$179.1&0.064&1.0&\dots\\
& C2 & 0.0284&0.0016&1.12&0.10&169.5&0.470&1.0&\dots\\
& C1 & 0.0171&0.0012&3.5&0.3&166.9&1.668&1.0&\dots\\
2011--07--15
& Core & 0.53&0.03&0.000&0.011&0.0&0.026&1.0&\dots\\
& C3 & 0.165&0.008&0.19&0.02&$-$174.5&0.082&1.0&\dots\\
& C2 & 0.0264&0.0014&1.15&0.08&170.5&0.437&1.0&\dots\\
& C1 & 0.0166&0.0010&3.5&0.2&166.2&1.769&1.0&\dots\\
2012--01--02
& Core & 0.43&0.02&0.000&0.014&0.0&0.027&1.0&\dots\\
& C3 & 0.133&0.007&0.20&0.03&$-$178.3&0.105&1.0&\dots\\
& C2 & 0.0264&0.0016&1.17&0.10&169.0&0.417&1.0&\dots\\
& C1 & 0.0174&0.0011&3.4&0.3&166.4&1.940&1.0&\dots\\
2012--03--27
& Core & 0.50&0.03&0.000&0.009&0.0&0.024&1.0&\dots\\
& C3 & 0.131&0.007&0.20&0.02&179.8&0.126&1.0&\dots\\
& C2 & 0.0257&0.0014&1.18&0.07&168.7&0.398&1.0&\dots\\
& C1 & 0.0170&0.0010&3.6&0.2&167.1&2.144&1.0&\dots\\
2012--11--11
& Core & 0.317&0.016&0.000&0.022&0.0&0.045&1.0&\dots\\
& C3 & 0.170&0.009&0.18&0.04&$-$167.8&0.142&1.0&\dots\\
& C2 & 0.0213&0.0024&1.23&0.18&170.4&0.447&1.0&\dots\\
& C1 & 0.0126&0.0015&3.6&0.4&168.5&1.893&1.0&\dots\\
2013--07--08
& Core & 0.352&0.018&0.000&0.010&0.0&0.022&1.0&\dots\\
& C3 & 0.084&0.004&0.26&0.03&$-$167.8&0.192&1.0&\dots\\
& C2 & 0.0133&0.0008&1.30&0.10&171.3&0.469&1.0&\dots\\
& C1 & 0.0125&0.0007&3.6&0.2&167.1&1.998&1.0&\dots\\
\hline
\end{tabular}
\end{minipage}
\end{table}

\begin{table}
\begin{center}
 \begin{minipage}{168mm}
  \contcaption{}
  \begin{tabular}{@{}lcr@{$\pm$}lr@{$\pm$}lcccc@{}}
\hline
Date & Name & \multicolumn{2}{c}{Flux density} & \multicolumn{2}{c}{Distance} & P.A. & Major & Ratio & Major P.A.\\
 &  & \multicolumn{2}{c}{(Jy)} & \multicolumn{2}{c}{(mas)} &($^\circ$)& (mas) & & ($^\circ$) \\
(1)& (2) & \multicolumn{2}{c}{(3)} & \multicolumn{2}{c}{(4)} &(5)& (6) &(7) & (8) \\
\hline
\multicolumn{10}{c}{23.4~GHz}\\
\hline
2010--05--24
& Core & 0.252&0.013&0.000&0.007&0.0&0.016&1.0&\dots\\
& C3 & 0.082&0.004&0.097&0.013&$-$170.9&0.047&1.0&\dots\\
& C2 & 0.0135&0.0007&1.01&0.06&172.7&0.497&1.0&\dots\\
& C1 & 0.0119&0.0006&3.29&0.14&166.3&1.460&1.0&\dots\\
2010--07--9
& Core & 0.296&0.015&0.000&0.006&0.0&0.018&1.0&\dots\\
& C3 & 0.077&0.004&0.105&0.013&$-$170.7&0.050&1.0&\dots\\
& C2 & 0.0148&0.0009&0.99&0.07&173.4&0.494&1.0&\dots\\
& C1 & 0.0169&0.0009&3.55&0.16&167.8&2.037&1.0&\dots\\
2010--08--28
& Core & 0.385&0.019&0.000&0.005&0.0&0.015&1.0&\dots\\
& C3 & 0.074&0.004&0.113&0.012&$-$173.1&0.046&1.0&\dots\\
& C2 & 0.0159&0.0009&1.03&0.06&172.3&0.472&1.0&\dots\\
& C1 & 0.0153&0.0009&3.11&0.17&167.1&1.795&1.0&\dots\\
2010--10--18
& Core & 0.47&0.02&0.000&0.003&0.0&0.016&1.0&\dots\\
& C3 & 0.129&0.006&0.111&0.007&$-$169.1&0.049&1.0&\dots\\
& C2 & 0.0166&0.0009&1.05&0.05&173.2&0.505&1.0&\dots\\
& C1 & 0.0118&0.0007&3.34&0.15&167.6&1.596&1.0&\dots\\
\hline
\multicolumn{10}{c}{43.2~GHz}\\
\hline
2010--05--24
& Core & 0.319&0.016&0.000&0.005&0.0&0.020&1.0&\dots\\
& C3 & 0.097&0.005&0.087&0.009&$-$172.4&0.029&1.0&\dots\\
& C2 & 0.0096&0.0007&1.09&0.08&177.5&0.478&1.0&\dots\\
2010--07--9
& Core & 0.4298&0.0215&0.000&0.005&0.0&0.015&1.0&\dots\\
& C3 & 0.0590&0.0031&0.107&0.014&$-$169.2&0.039&1.0&\dots\\
& C2 & 0.0054&0.0007&1.16&0.07&172.2&0.164&1.0&\dots\\
2010--08--28
& Core & 0.5858&0.0293&0.000&0.003&0.0&0.021&1.0&\dots\\
& C3 & 0.0701&0.0035&0.109&0.010&$-$169.6&0.043&1.0&\dots\\
& C2 & 0.0155&0.0010&1.18&0.10&172.0&0.716&1.0&\dots\\
2010--10--18
& Core & 0.6929&0.0346&0.000&0.002&0.0&0.016&1.0&\dots\\
& C3 & 0.1054&0.0053&0.116&0.007&$-$169.7&0.050&1.0&\dots\\
& C2 & 0.0115&0.0007&0.96&0.09&170.7&0.837&1.0&\dots\\
\hline
\end{tabular}
\end{minipage}
\end{center}
\end{table}

\bsp
\label{lastpage}
\end{document}